\documentclass[10pt,journal,compsoc]{IEEEtran}
\ifCLASSOPTIONcompsoc
  \usepackage[nocompress]{cite}
\else
  \usepackage{cite}
\fi
\ifCLASSINFOpdf
   \usepackage[pdftex]{graphicx}
\else
\fi
\usepackage{bm}
\usepackage{amssymb}
\usepackage{amsfonts}
\usepackage[margin=2.5cm]{geometry}
\usepackage{graphicx}
\usepackage{multirow}
\usepackage{url}
\usepackage{hyperref}
\usepackage[hyphenbreaks]{breakurl}
\usepackage[table]{xcolor}
\usepackage{soul}
\usepackage{amsmath}
\usepackage{algorithm}
\usepackage{algorithmic}

\graphicspath{{./}{images/}}
\usepackage{array}  
\newcolumntype{H}{>{\setbox0=\hbox\bgroup}c<{\egroup}@{}} % Hiding entire column of a table without removing it.

\begin{document}

%\title{Localization of Quantum walks in complex networks}
\title{Multilevel Digital Contact Tracing}
\author{Gautam Mahapatra, Priodyuti Pradhan, Abhinandan Khan, Sanjit Kumar Setua, Rajat Kumar Pal and  Ayush Rathor %\today %~\IEEEmembership{Member,~IEEE,}
\IEEEcompsocitemizethanks{\IEEEcompsocthanksitem Gautam Mahapatra is associated with Department of Computer Science, Asutosh College, University of Calcutta, Kolkata, India. E-mail: gsp2ster@gmail.com

\IEEEcompsocthanksitem Priodyuti Pradhan is associated with the networks.ai Lab, Department Computer Science \& Engineering, Indian Institute of Information Technology Raichur, Karnataka - 584135, India. E-mail: priodyutipradhan@gmail.com

\IEEEcompsocthanksitem Abhinandan Khan is associated with the Department of Computer Science \& Engineering, University of Calcutta, India. E-mail: khan.abhinandan@gmail.com

\IEEEcompsocthanksitem Sanjit Kumar Setua is associated with the Department of Computer Science \& Engineering, University of Calcutta, India. E-mail: sksetua@gmail.com

\IEEEcompsocthanksitem Rajat Kumar Pal is associated with the Department of Computer Science \& Engineering, University of Calcutta, India. E-mail: palrajatk@gmail.com

\IEEEcompsocthanksitem Ayush Rathor is associated with the networks.ai Lab, Department Computer Science \& Engineering, Indian Institute of Information Technology Raichur, Karnataka - 584135, India. E-mail: ayushrathore457@gmail.com
}% <-this % stops an unwanted space

}

% The paper headers
\markboth{Journal of \LaTeX\ Class Files,~Vol.~14, No.~8, \today}%
{Shell \MakeLowercase{\textit{et al.}}: Bare Demo of IEEEtran.cls for Computer Society Journals}

\IEEEtitleabstractindextext{%
\begin{abstract}
Digital contact tracing plays a crucial role in alleviating an outbreak, and designing multilevel digital contact tracing for a country is an open problem due to the analysis of large volumes of temporal contact data. We develop a multilevel digital contact tracing framework that constructs dynamic contact graphs from the proximity contact data. Prominently, we introduce the edge label of the contact graph as a binary circular contact queue, which holds the temporal social interactions during the incubation period. After that, our algorithm prepares the direct and indirect (multilevel) contact list for a given set of infected persons from the contact graph. Finally, the algorithm constructs the infection pathways for the trace list. We implement the framework and validate the contact tracing process with synthetic and real-world data sets. In addition, analysis reveals that for COVID-19 close contact parameters, the framework takes reasonable space and time to create the infection pathways. Our framework can apply to any epidemic spreading by changing the algorithm's parameters.
\end{abstract}
%under Public Health Authorities' control. Our framework 
% Note that keywords are not normally used for peer review papers.
}

% make the title area
\maketitle

\IEEEdisplaynontitleabstractindextext
\IEEEpeerreviewmaketitle
\IEEEraisesectionheading{\section{Introduction}\label{sec:introduction}}

\IEEEPARstart{I}nfectious diseases caused by microscopic germs such as bacteria or viruses are called the infection elements that get into the human body and cause health problems. Infectious diseases that spread from person to person on natural social contact are said to be contagious. Within a definite past period, the procedure of identifying a list of persons who come directly or indirectly in social contact for a given set of persons is called \emph{contact tracing}. In case of contagious disease, if there is a list of infected persons, then contact tracer results in a possible suspected list that may carry the infection-spreading germs. 
%Hypothetically speaking, we can enclose the infection-spreading germs at a particular time by successfully generating accurate contact trace results for all infected persons.
We can enclose the infection and suspected individuals for a particular time by successfully generating accurate contact trace for all infected persons.
We can monitor trace listed individuals for a predefined period (say $D$), which we call the incubation period (Ebola $D=21$, Influenza $D=5$, and for COVID-19, $D=14$ \cite{contact_trace_Ebola_2014, contact_trace_H1N1_2009}). Within this incubation period, we may get confirmation of disease-specific symptoms, and lab test reports for the enclosed suspected list. Based on that, we can provide proper care and medical treatment for the listed infected individuals.
%so that, along with other non-infected members, they are returning to society as normal members. 
Different studies show that \emph{contact tracing} and case isolation has become one of the essential non-pharmaceutical methods to mitigate the new outbreak of infectious disease \cite{digital_contact_2023, contact_trace_Ebola_2014,hellewell2020feasibility, future_covid} and authorities use this as a common and successful step to control several infectious diseases like Zika, HIV, influenza, Ebola, COVID-19 viruses  \cite{world2020coronavirus, de2020wetrace}. However, in practice, due to several real-life situations, successful implementations and accuracy in contact tracing are not achievable, so the \emph{contact tracing} remains an ever-challenging problem. 

Contact tracing aims to find people early in their course of disease so they can be offered treatment and have a higher chance of survival. The goal of contact tracing is to reduce a disease's effective reproductive number ($R_0$) by identifying people who have been exposed to the virus through close contact with one or more infected persons directly or indirectly and listing them for immediate isolation \cite{Effectiveness_digital_contact_2022, DCT_2022, DCT_network_theory_2022}. It has been reported that $46\%$ contribution to $R_0$ comes from the presymptomatic individual (before showing symptoms) for COVID-19 \cite{ferretti2020quantifying}. Presymptomatic individuals are infectious before they develop symptoms, typically during the incubation period of the disease. Presymptomatic cases are challenging to detect, as individuals are not yet showing symptoms. Contact tracing and testing of close contacts are essential for identifying presymptomatic cases (Table  1, Appendix). Hence, highly effective direct and indirect contact tracing is a mandatory task that plays a crucial role in the detection of early infectious germs carriers and reduces the peak burden on the healthcare system. 

Due to higher uncertainties in social relations, manual contact tracing to find the contact structure is very complicated for the Public Health Authority (PHA) \cite{keeling2020efficacy}. Proximity sensors-enabled smartphones (Bluetooth, Global Positioning System, WiFi, Ultrasound signaling, etc.) help to implement digital contact tracing to overcome the problem in manual contact tracing \cite{importance_digital_contact, nature_digital_contact, science_digital_contact}. The TraceTogether of Singapore Government \cite{TraceTogether2020}, COVID-Watch of Stanford University \cite{CovidWatch2020CT, CovidWatch2020WhitePaper}, PACT of MIT \cite{rivest2020pact}, Exposure Notification App of Apple-Google \cite{AppleGoogle2020CT}, Arogya Setu of MoHFA, Government of India \cite{aarogyasetu2020} are all recent success stories of digital contact tracing app that are effectively used to mitigate the COVID-19 pandemic \cite{oliver2020mobile, ferretti2020quantifying, hernandez2020evaluating, amit2020mass, maxmen2019can, Turing2020}. During the pandemic emergency, the PHA conducts laboratory testing for infection detection. For positive cases, PHA arranges contact tracing tasks along with medical treatment, isolation-quarantine, and many other protective measures. The newly developed computer software-apps-based digital contact tracing systems reduce human efforts substantially, but like long-existing manual contact tracing systems, these are considering only the direct social contacts of the infected person. However, carriers of infectious germs may not show any symptoms of infection for a long time, even for the whole incubation period (presymptomatic and asymptomatic nature). Still, they can spread contagious infection through social contact. Therefore, multilevel, i.e., both direct and indirect infection spreading may happen, which affects the $R_0$ \cite{vietnam2020, recursive_contact_tracing_2021, proactive_contact_tracing_2023}.
Our study considers not only direct contacts of the confirmed cases but also the contacts of those contacts and so on, which we call multilevel digital contact tracing \cite{WHOwarning2020}. It helps to identify potential transmission chains more comprehensively and quickly, thus mitigating outbreaks more effectively (details in Appendix sec. 1). 

%We can identify benefits of the multilevel contact tracing as -- {\bf a) greater coverage.} It extends the reach of tracing efforts beyond just the immediate contacts of confirmed cases, ensuring a more thorough identification of potential transmission chains. {\bf b) early detection.} By tracing contacts beyond the primary contacts, multilevel contact tracing can detect potential cases earlier in the transmission chain, allowing for prompt isolation and treatment, thereby preventing further spread. {\bf c) early warning.} Identifying contacts beyond immediate interactions can provide early warning signs of potential outbreaks, allowing for timely intervention. {\bf d) enhanced control.} Multilevel tracing generates rich data that can be used for detailed analysis and modeling of disease spread, leading to more informed decision-making. By tracing multiple levels of contact, public health authorities (PHA) gain a deeper understanding of transmission dynamics, allowing for more targeted and effective interventions. It also helps to optimize resource allocation by focusing on high-risk contacts and locations. Overall, multilevel contact tracing can significantly enhance the effectiveness of public health interventions in managing and controlling infectious diseases.

This article provides a framework to automate multilevel digital contact tracing. Our algorithm processes the contact data and dynamically evolves a close contact graph. The nodes in the close contact graph are the individual users, and an edge is included if an interaction between a pair of people exists. Importantly, to store the temporal close contact information, we introduce an edge label between a pair of individuals in the contact graph as a fixed size binary circular contact queue. In other words, during a pandemic, the contact graph stores the latest $D$ days of continuous close contact data in a discrete form inside circular contact queues.
Finally, the system prepares the contact list for a given infected person from the contact graph. Our algorithm prepares the direct and indirect (multilevel) contact lists of the infected persons and prepares the infection pathways. The system's salient feature is that it automatically removes the inactive edge (contact) when the $D$ days over and updates the contact queues. We provide an analytical approach for the completeness of the algorithm and validate it using synthetic and empirical data sets. As a case study, our analysis reveals that for COVID-19 contact trace parameters, storing the contact graph for $14$ days for $10^{6}$ users takes $5$ GB of memory space, and the preparation of the contact list for a given set of infected persons depends on the size of the infected list. Our algorithm is simple and easy to implement. We expect it to be an attractive choice to deploy in the application of digital contact traces in real-world pandemic situations.

\section{Methodology and Results}\label{methods}
For the digital contact tracing system, a population denoted as $V$ is a set of individuals under a particular Public Health Authority (PHA) who use their smartphones. For human nature of social interactions, any two individuals $P, P^{'}\in V$, we define $P$ is in { \em close contact} of $P^{'}$ (or $P^{'}$ is in { \em close contact} of $P$), if during last $D$ days, $P$ and $P^{'}$ are in proximity of contact within $d$ meters of distance for at least $\tau$ minutes  \cite{bar2020science}. For a particular contagious disease, $d$ (proximity distance), $\tau$ (proximity duration), and $D$ (incubation period) are parameters of {\em close contact}, respectively. 

\begin{figure}[tbh]
    \centering
    \includegraphics[width=3.1in, height=1.9in]{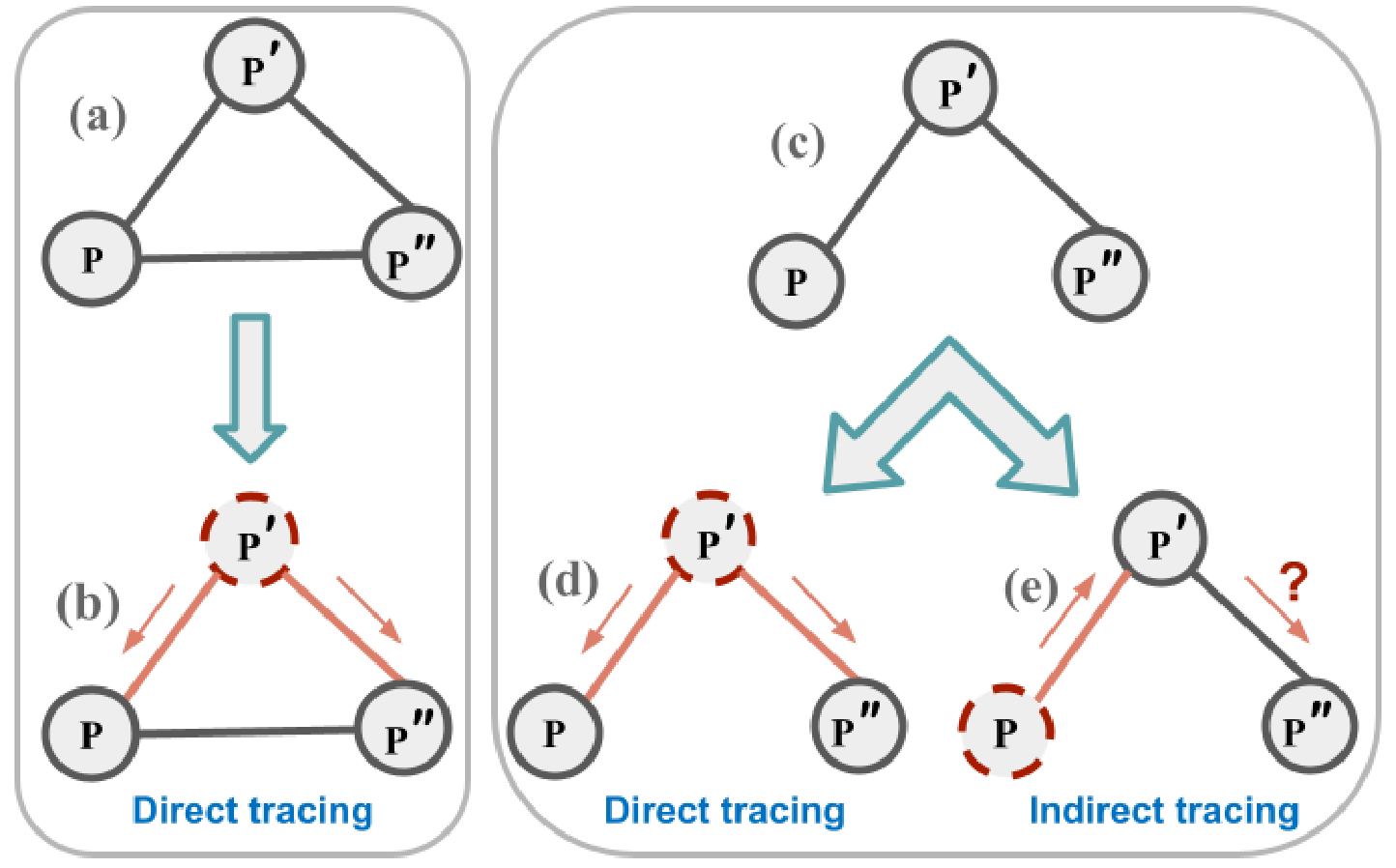}
    \caption{Illustration of direct and Indirect contact tracing. The dotted circle represents the infected person.}
    \label{fig:DirectIndirect}
\end{figure}
Importantly, when $P$ and $P^{'}$ come in close contact, they have no disease symptoms but may carry infection elements. Even they do not know each other. On the other hand, they also meet several times in a single day; thus, several close contacts between $P$ and $P^{'}$ in D days. If anyone has the infection, the symptoms will appear within the next $D$ days. Therefore, to find the contact list for an infected person, it is essential to hold the temporal information of all the close contacts for the next $D$ days. 

To illustrate close contact, we consider three individuals ($P$, $P^{'}$, and $P^{''}$) and their possible close contacts (Fig. \ref{fig:DirectIndirect}). One can observe that if $P^{'}$ is detected as an infected person ($\mathcal{I}=\{P^{'}\}$) within $D$ days of the close contact then $P$ and $P^{''}$ belongs to the {\em trace list} denoted as $\Gamma_{P^{'}}=\{P, P^{''}\}$ of $P^{’}$ (Fig. \ref{fig:DirectIndirect}(b) and (d)). Here, $\{P, P^{''}\}$ are represented as {\em direct} contact of an infected person $P^{'}$. We show another complex situation, which is also very important to trace and help to reduce the spread of the disease \cite{vietnam2020}. Suppose in Fig. \ref{fig:DirectIndirect} (c, e), $P$ is detected as an infected person, then $P^{'}$ should be in the direct contact trace list of $P$ i.e., $\Gamma_{P}=\{P^{'}\}$. However, whether $P^{''}$ should be in the close contact list of $P$ or not? To decide, we should analyze the temporal information between $(P, P^{'})$ and  $(P^{'}, P^{''})$ for the last $D$ days. Here, {\em $P^{''}$ will be in the trace list of $P$, if $P^{''}$ has a close contact with $P^{'}$ during or after the close contact between $P$ and $P^{'}$ within $D$ days}. We refer $P^{''}$ as {\em indirect} contact for $P$. The challenge is that we have to store a large volume of proximity contact data for a certain period (incubation period), and recursive temporal analysis is required to find the contact list. 

Formally, a contact tracing system for a contagious disease must store all {\em close contact} information of the whole population for the last $D$ days. Thus, at any time for a given set of infected individuals ($\mathcal{I}$), the system can determine the trace list ($\Gamma$). For simplicity, we assume once an individual is detected as a laboratory-confirmed infected person, immediately contact tracing is carried out, and the infected, as well as all contacted individuals, are isolated so that they do not contact further. 

There are two major components of the contact tracing system: {\bf module 1} (mobile app) -- data acquisition by smartphones -- capturing of proximity social contacts through Bluetooth enabled devices, subsequently conversion into discrete form and send to the PHA, {\bf module 2} -- computation on PHA side: a) analyze the discrete close contact data  (b) uses an efficient {\em storage representation} to hold all such close contact data for the last $D$ days period, (c) design an {\em effective contact tracing algorithm} working on such voluminous data to generate contact trace results within a reasonable time, and can be used to show the {\em infection pathways}. This article only focuses on the algorithm and implementation of {\bf (module 2)}.

\subsection{Contact graph and Digitization of Close Contacts}\label{digitization_close_contact}
\begin{figure*}[tbh]
    \centering
\includegraphics[width=7in, height=3.1in]{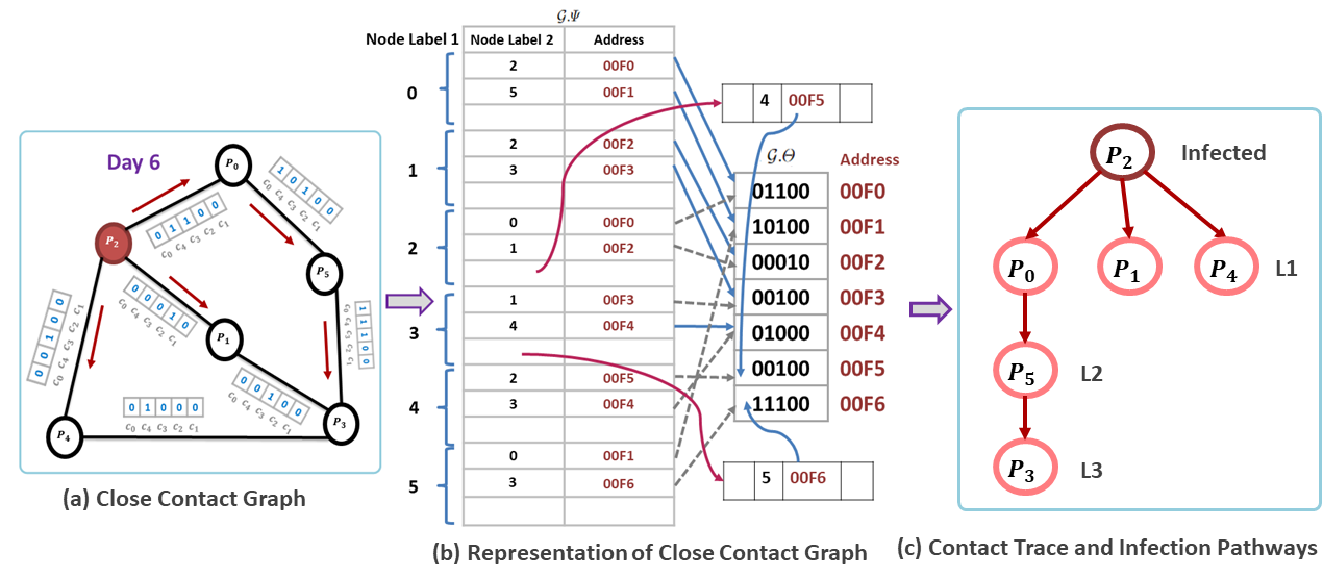}
\caption{(a) A close contact graph ($\mathcal{G}$) with $N=6$ nodes where circular contact queues (CCQ) are on the edges and $P_2$ is marked as infected (Fig. \ref{ExampleContactGraph}(Day 6)). (b) Contact graph representation using the array ($\mathcal{G}.\Psi$) with two fields (Node label $2$ or UserID and Address to store CCQ). Further, $\mathcal{G}.\Theta$ stores all CCQs. Here, we consider $q=2$; thus, for $P_2$ and $P_3$, adjacent information will also be stored in the link list. (c) From $\mathcal{G}$, during the contact tracing process, we find multilevel trace list $\Gamma_{P_2} = \{P_{0}^{1}, P_{1}^{1}, P_{4}^{1}, P_{5}^{2}, P_{3}^{3}\}$. Also, one can very easily construct the infection transmission pathways by storing the directed edges, $\chi_{P_2} = \{(P_2, P_0),(P_2, P_1),(P_2, P_4),(P_0, P_5),(P_5, P_3) \}$ where $(P_i, P_j)$ represents $P_i$ transmits infection to $P_j$.}
\label{contact_graph_representation}
\end{figure*}
\begin{figure*}[tbh]
    \centering
\includegraphics[width=7in, height=3.6in]{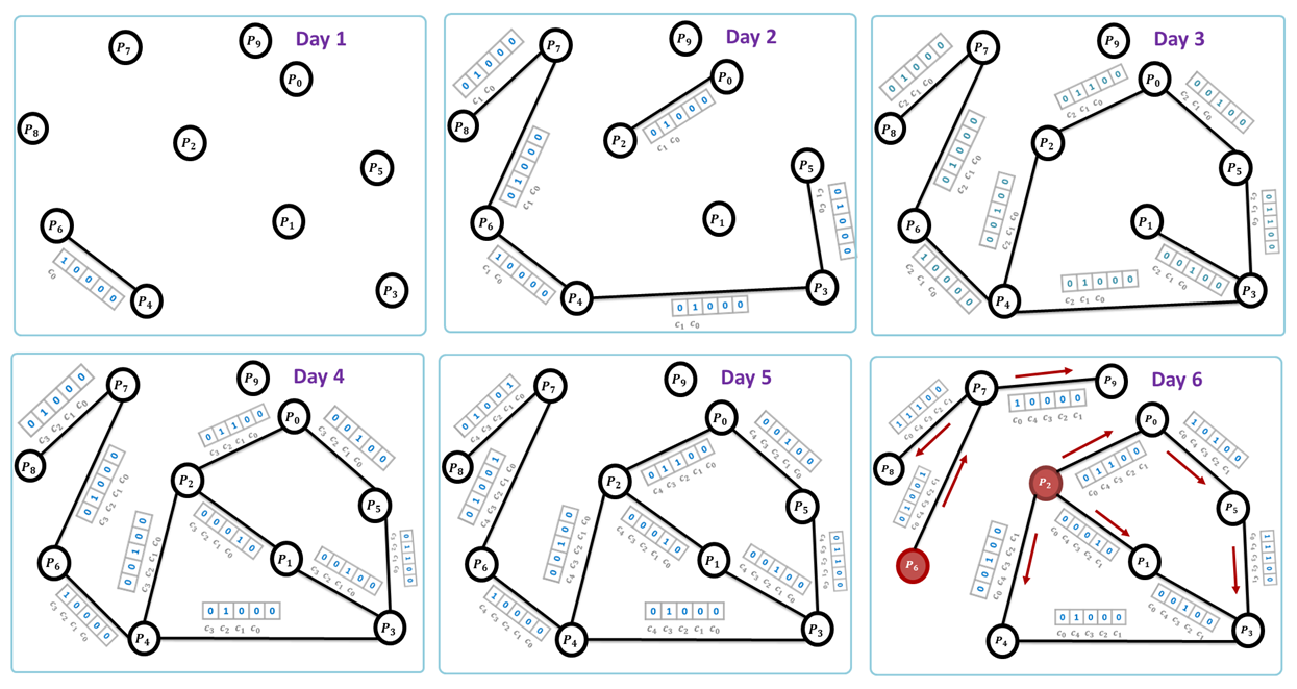}
\caption{Dynamic evolution of close contact graph ($\mathcal{G}$) for six days annotated with the circular contact queues ($\bm{c}$). We consider $V = \{P_0,P_1,\dots,P_9\}$, $D=5$ days, $\tau=1$ day and thus $N=|V|=10$ and size of $\bm{c}$ as $n=5$. Here, $c_0$ points to the latest, and $c_4$ points to the oldest day close contact information. Let's say initially, on the day $1$, there is a close contact between $P_4$ and $P_6$; thus, there will be an edge and left most bit position in $\bm{c}_{(P_4, P_6)}$ set to one and pointed by $c_0$. In the next day there is another five contacts ($\{(P_0,P_2),(P_3,P_4),(P_3,P_5),(P_6,P_7), (P_7,P_8)\}$) but no contact between $P_4$ and $P_6$. Thus, for day $2$, $c_0$ position is zero in $\bm{c}_{(P_4, P_6)}$ and one for the rest of the contact queues where $c_1$ points to day $1$ data. In the same way, the contact graph dynamically updates as contacts are received from individual users. After the incubation period ($5$ days) is over, on the $6$th day, we are not required to keep the day $1$ contact data. One can notice that there is an edge between $P_4$ and $P_6$ during the $5$ days; however, at $6th$ day, they have no close contact; thus, the algorithm removes the edge $(P_4, P_6)$, updates $\mathcal{G}$ and $c_0$ start pointing to the left side cell and the process repeats. On the other hand, one can also observe that during the $5$ days, there is no close contact between $P_7$ and $P_9$. However, on $6^{th}$ day, there is a close contact between $P_7$ and $P_9$; thus, the algorithm sets one bit in $\bm{c}_{(P_7, P_9)}$ pointed by $c_0$. Further, we assume on Day 6, two persons are detected as infected ($P_2$ and $P_6$).}
\label{ExampleContactGraph}
\end{figure*}

We consider a country subdivided into several zones, each with one PHA that maintains all computations of the multilevel digital contact tracing process. While PHA receives the discretized proximity contacts $(Time,P,P^{'})$, immediately records in a dynamic contact graph structure, $\mathcal{G}=(V, E,\bm{c}_{(P, P^{'})})$. Here, vertices of $\mathcal{G}$ are the smartphone-enabled individuals ($V$) where $N=|V|$. If there is any {\em close contact} between two individuals, i.e., vertices $P, P^{'}\in V$, then we have an undirected edge, $e_{i}=(P, P^{'},\bm{c}_{(P, P^{'})})$  in $\mathcal{G}$, and store the temporal close contact information as bits inside a {\em FIFO circular contact queue} (Fig. \ref{contact_graph_representation}(a)), $\bm{c}_{(P, P^{'})}=(c_{n-1},\ldots c_1,c_0)$ where $c_{n-1}$ is the oldest day close contact and $c_0$ is the latest day contact (Appendix sec. 3). In the digitalization of social contacts, we use one bit for each $\tau$ minutes of {\em close contacts}. Hence, we require $n=\lceil \frac{24\times 60 \times D}{\tau}\rceil$ number of bits to capture all {\em close contacts} during the $D$ days incubation period. For a particular disease parameters, $D$ and $\tau$ are fixed, thus  $\bm{c}_{(P,P^{'})}$ has a fixed length ($n$). By default, all $n$ bits are \textsc{zero}, and a bit is set to \textsc{one} if $P$ and $P^{'}$ come close to each other for $\tau$ minutes (Appendix Fig. A1). For instance, to identify every {\em close contacts} in the COVID-19 pandemic, we subdivide the $D=14$ days into $\tau=15$ minutes time slots, and hence the length of $\bm{c}$ is $n=1344$ bits. 
Importantly, as time advances to the first slot of $(D+1)^{th}$ day, then the oldest slot at $c_{n-1}$ exits to accommodate a new input slot at $c_{0}$ (Fig. \ref{ExampleContactGraph}). In other words, after $D$ days are over, the system automatically updates the queue and removes the edges. Hence, we use the circular queue to store the close contact information.

During the update of an edge and its associated circular contact queue (CCQ) in $\mathcal{G}$, individuals do not know whether they are infected. Hence, the edges of the contact graph are always undirected ($\bm{c}_{(P, P^{'})}=\bm{c}_{(P^{'}, P)}$). If any individual has any symptoms and performs a test in the laboratory, they can be detected as infected. Then, the tracing algorithm can trace the contacts for the infected person over the close contact data stored within the CCQ in $\mathcal{G}$ and find the trace list and infection pathways. 

For a sufficiently large population, due to social relationships, everyone is not coming in close contact with every other \cite{danon2013social}. Therefore, we can consider individuals meeting each other with a random distribution $P(q)$. It indicates that the average number ($\langle q\rangle$) of close contact relationships is tiny compared to $ N $, $\langle q\rangle \ll N$, and hence $\mathcal{G}$ is highly sparse. On the other hand, once the incubation periods are over, the existing edges will be removed from the contact graphs automatically; thus, $\mathcal{G}$ is dynamic. 

To store such a highly sparse and dynamic close contact graph, we use a modified version of the adjacency lists \cite{cormen2009introduction} with two components (Fig. \ref{contact_graph_representation}(b) and Fig. A2), adjacent file ($\mathcal{G}.\Psi$) and circular close contact queues file ($\mathcal{G}.\Theta$). The $\mathcal{G}.\Psi$ stores $q$ number of adjacent information for each user with two fields (\textit{UserID} and CCQ \textit{Address}) (Fig. \ref{contact_graph_representation}(a, b)). We know for a pair of individuals $P$ and $P^{'}$ there is an undirected edge having $\bm{c}_{(P,P^{'})}$ or $\bm{c}_{(P^{'},P)}$ as the edge label. Hence, we store $\bm{c}$ only once in $\mathcal{G}.\Theta$ at $A$ address position in $\mathcal{G}.\Psi$. For an edge $e=(P,P^{'},\bm{c}_{(P,P^{'})})$, $P^{'}$ will appear as one record $(P^{'}, A)$ under adjacency list of $P$ in $\mathcal{G}.\Psi$. Similarly, we keep another record $(P, A)$ under $P^{'}$. Moreover, for an individual, we maintain one extra slot in $\mathcal{G}.\Psi$ to move into the overflow area when the number of contacts exceeds $q$ contacts. Therefore, each person has $q+1$ slots and total $N(q+1)$ locations in $\mathcal{G}.\Psi$. Therefore, we can access $\mathcal{G}.\Psi$ as an array with constant access time. Here, algorithm uses $\mathcal{G}.\Psi[P\times(q+1)]$ to access the starting contact information of $P$. To maintain the overflow contact information, we use a link-list data structure \cite{cormen2009introduction}, which dynamically grows or shrinks as time progresses (Fig. \ref{contact_graph_representation}(b)). 

\begin{figure*}[tbh]
\begin{center}
\includegraphics[width=6.8in, height=4.5in]{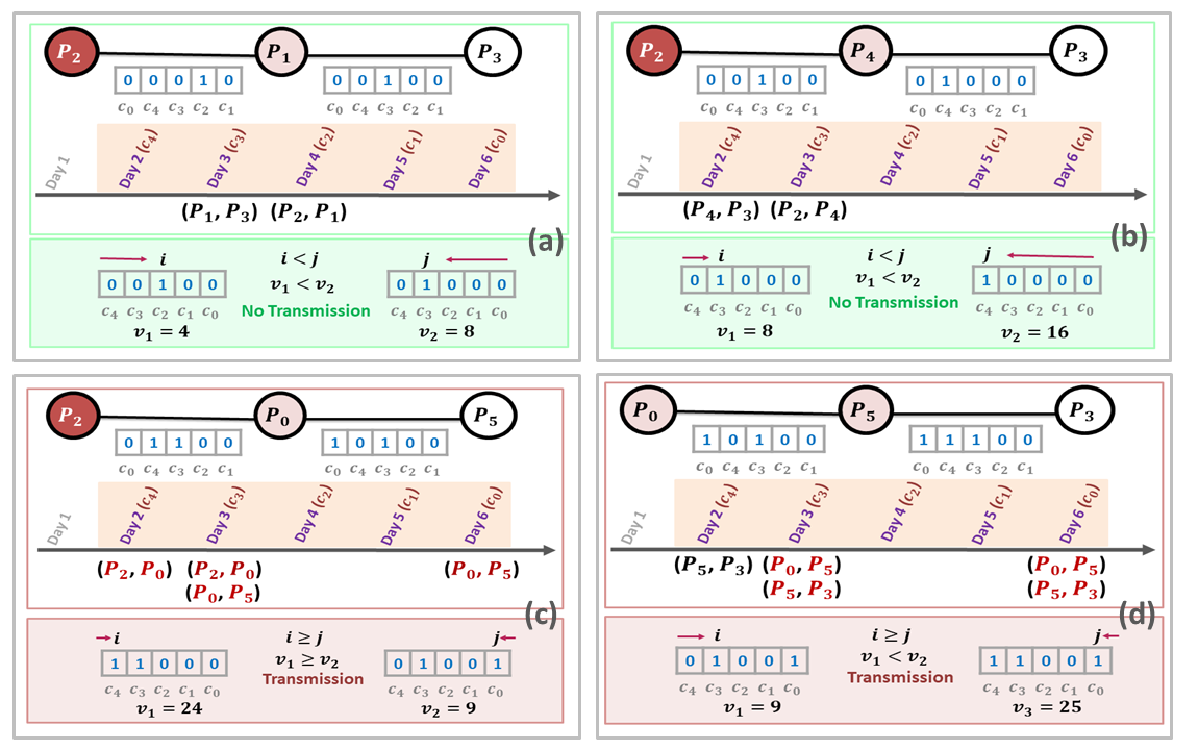}
\caption{Indirect contact tracing. We unfold the indirect contact paths where $P_2$ is the infected node in Fig. \ref{ExampleContactGraph}(Day 6). (a) $P_3$ is in the indirect contact list of $P_2$. We can observe from the circular contact queues (CCQs) and timeline that on day $3$, there is a contact between $P_1$ and $P_3$. After that, on day $4$, there is a contact between $P_2$ and $P_1$, and no further contacts exist. Hence, infection can not spread from $P_2$ to $P_3$ via $P_1$. We can detect it by the alignment of the CCQ where $c_0$ is the least significant bit and $c_{n-1}$ is the most significant bit. We can observe that $v_1<v_2$ and $i<j$ ensure no indirect transmission. (b-d) Similarly, we can explore and detect the other paths.}
\label{indirect_trace}
\end{center}
\end{figure*} 

\subsection{Multilevel Contact 
Tracing}\label{contact_trace_example}

We require daily close contact data for a population to test the multilevel contact tracing algorithm. It is very difficult to get enough real-world close contact data to test the tracing algorithm. Therefore, we generate close contacts as synthetic data for our experiment (sec. \ref{synthetic_data}). On initialization, the PHA creates an empty $\mathcal{G}$ along with $\mathcal{I}=\{\}$, $\Gamma = \{\}$ and having population $V = \{P_0, P_1,\dots, P_{N-1}\}$. For a given infected list $\mathcal{I}$, we can easily generate all the {\em direct contact} trace members from the adjacent information of the contact graph ($\mathcal{G}$). However, for the {\em indirect contact} list, we need to analyze the CCQs. Here, we illustrate the contact tracing process with an example. Fig. \ref{ExampleContactGraph} is a scenario of $\mathcal{G}$ with ten individuals $V = \{P_0,P_1,\dots,P_9\}$, thus $N=10$. For simplicity, we consider $D=5$ days, $\tau=1$ day, thus $n=5$. We show the evolution of $\mathcal{G}$ for six days with the associated CCQ on the edges (Fig. \ref{ExampleContactGraph}). Let us assume up to $5th$ day, $\mathcal{I}=\{\}$, hence, $\Gamma = \{\}$ (Fig. \ref{ExampleContactGraph}). On the $6th$ day $P_2$ and $P_6$ (marked as red in Fig. \ref{ExampleContactGraph}) are identified as infected ($\mathcal{I}=\{P_2, P_6\}$) from the laboratory test; hence, the PHA initiates the contact tracing process on $\mathcal{G}$ and finds the contact trace list, $\Gamma=\Gamma_{P_2} \cup \Gamma_{P_6}$. 

\subsubsection{Direct Tracing}
We can observe that on day $6$, $P_0$, $P_1$ and $P_4$ are in direct close contact of $P_2$ in latest $D$ days as there are edges and corresponding circular contact queues are nonzero ($P_2 \rightsquigarrow P_0$, $P_2 \rightsquigarrow P_1$ and $P_2 \rightsquigarrow P_4$) from $\mathcal{G}$ (Fig. \ref{ExampleContactGraph}). Hence, from $\mathcal{G}$, $\Gamma_{P_2} = \{P_{0}^{1}, P_{1}^{1}, P_{4}^{1}\}$ and $\Gamma_{P_6} = \{P_{7}^{1}\}$ are the direct or first-level contact trace list of $P_2$ and $P_6$  where $P_{x}^{y}$ represents $P_x$ appears in $y$-level.   

\subsubsection{Indirect Tracing}\label{indirect_contacts}
To prepare the second-level (indirect) close contact list of $P_2$ on day $6$, one can observe that $P_3$ and $P_5$ are possible candidates (Fig. \ref{ExampleContactGraph}). However, we can not confirm this by observing only the presence of edges and nonzero CCQs. Now, we need to analyze the associated CCQs. Remarkably, the listing and pruning of second and higher-level contact lists can be prepared through numerical operations on the equivalent decimal integer values corresponding to the aligned binary CCQ where $c_{0}$ as Least-significant bit (LSB) and $c_{n-1}$ as Most-significant bit (MSB) of the binary number. In other words, $c_0$ points to the latest day, and $c_{n-1}$ points to the oldest day close contacts (Appendix sec. 3).

From Figs. \ref{ExampleContactGraph}(Day 6) and \ref{indirect_trace}(a, b), there are two ways $P_3$ can get the infection indirectly either from $P_1$ or from $P_4$. From the CCQs, $\bm{c}_{(P_1, P_2)}=00010$, $\bm{c}_{(P_1, P_3)}=00100$, and if we look into the timeline in Fig. \ref{indirect_trace}(a, b), one can observe that $P_1$ meets $P_3$ on day $3$ and $P_1$ meets $P_2$ on day $4$, thus on sixth day if $P_2$ is detected as infected, the infection does not pass from $P_2$ to $P_3$ via $P_1$. We capture the phenomena from the decimal values of the corresponding aligned CCQ ($v_1=val(\bm{c}_1)=val(\bm{c}_{(P_1, P_2)})=00100=4$, $v_2=val(\bm{c}_2)=val(\bm{c}_{(P_1, P_3)})=01000=8$) and observe that $v_1 < v_2$ and $i<j$ (Lemma 2, Appendix) where $i$ is the oldest bit which is one in $\bm{c}_1$ and $j$ is the latest bit which is one in $\bm{c}_2$ with $0 \leq i,j \leq n-1$ (Fig. \ref{indirect_trace}(a)). Similarly, from $\bm{c}_{(P_2, P_4)}$ and $\bm{c}_{(P_3, P_4)}$, we can also infer that $P_4$ does not pass infection to $P_3$ and the decimal values also ensure it, i.e., $v_1=val(\bm{c}_{(P_2, P_4)})=8$ and $v_2=val(\bm{c}_{(P_3, P_4)})=16$, thus, again $v_1 < v_2$ and $i<j$ (Fig. \ref{indirect_trace}(b)). Therefore, the inclusion of $P_3$ into the second level contact list is not required ($P_2 \rightsquigarrow P_1\not \rightsquigarrow P_3$ and $P_2 \rightsquigarrow P_4 \not \rightsquigarrow P_3$). Further, there is one way $P_5$ may get infection from $P_2$ via $P_0$ (Figs. \ref{ExampleContactGraph} and \ref{indirect_trace}(c)). From $\bm{c}_{(P_0, P_2)}$ and $\bm{c}_{(P_0, P_5)}$ one can observe that two-bit patterns imply that at day $2$, both $P_0$ and $P_2$ comes close to each other and have close contact, but $P_0$ and $P_5$ do not have any close contact. However, on the day $3$ in between ($P_0$, $P_2$) and ($P_0$, $P_5$) have close contacts, indicating the three are all together. Hence, infection can transmit from  $P_2$ to $P_5$ via $P_0$ ($P_2 \rightsquigarrow P_0\rightsquigarrow P_5$). We can extract the decimal integer values of $v_1=val(\bm{c}_{(P_0, P_2)})=11000=24$ and $v_2=val(\bm{c}_{(P_0, P_5)})=01001=9$ at day $6$ (Figs. (\ref{ExampleContactGraph}) and \ref{indirect_trace}(c)). Here, $v_1 \geq v_2$ and $i \geq j$ (Lemma 1, Appendix) says at least one contact between $P_0$ and $P_2$ occurred earlier than the contact between $P_0$ and $P_5$. Hence, $P_5$ can get infection from $P_2$ via $P_0$ and thus be included in the second-level contact list.

Now, we test for the third-level contact of $P_2$. We already know $P_2 \rightsquigarrow P_0 \rightsquigarrow P_5$ is \textsc{true}. We have to check whether $P_3$ will get the infection via $P_5$. $P_3$ will be included in the third level trace list if one close contact exists between $(P_3, P_5)$ after (or same day) a close contact between $(P_0, P_5)$. From $\bm{c}_{(P_0,P_5)}$, $\bm{c}_{(P_3,P_5)}$, and time line we can observe that in day $3$, $P_0$, $P_3$ and $P_5$ are close to each other. Here, $v_{1}=val(\bm{c}_{(P_0,P_5)})=01001=9$ and $v_{2}=val(\bm{c}_{(P_3,P_5)})=11001=25$ (Fig. \ref{indirect_trace}(d)). In this case, although $v_1<v_2$, $i \geq j$ (Lemma 2, Appendix) infers that infection may transmit because by analyzing the timeline, oldest ($i$) and latest ($j$) bit position of $\bm{c}_1$ and $\bm{c}_2$, we can understand the situation (Fig. \ref{indirect_trace}) (d). Therefore, 
$P_3$ gets infection from $P_2$ via $P_0$ and $P_5$. Hence, $P_3$ will be included in the third-level contact list (Fig. \ref{ExampleContactGraph}). No further processing is possible and tracing stops with $\Gamma_{P_2} = \{P_{0}^{1}, P_{1}^{1}, P_{4}^{1}, P_{5}^{2}, P_{3}^{3}\}$, and similarly $\Gamma_{P_6} = \{P_{7}^{1}, P_{8}^{2}, P_{9}^{2}\}$. Fig. \ref{contact_graph_representation}(c) shows the multilevel contact list and infection pathways tree for the larger component in Fig. \ref{ExampleContactGraph}(Day 6). Our contact tracing algorithm takes time $\mathcal{O}(q^{L}|\mathcal{I}^{'}|)$ where $q$ is the average number of contacted persons, $|\mathcal{I}^{'}|$ is the number of infected person for a particular day and $L$ is the maximum level for which we want to trace (sec. \ref{time_complx}). The storage space required for storing the contact graph for the incubation periods is $\mathcal{O}(N \log N)$ bits (sec. \ref{space_complx}). Note that we consider $\tau=1$ day for simplicity. Still, we can easily understand the more general situation when $\tau$ in minutes. The following section summarizes the contact tracing algorithm on the contact graph. 

\subsection{Multilevel Contact Tracing Algorithm}

The close contact graph ($\mathcal{G}$) maintains the digital form of the social contact dynamics for a particular geopolitical area under a PHA for the last $D$ days. We assume that at any time, the system has a contact trace list ($\Gamma$) for a given set of infected persons ($\mathcal{I}$). The system updates these lists using Algorithm 1 (Appendix) for incoming situations. Suppose at any time, one or more individuals are detected as newly infected persons and kept in $\mathcal{I}^{'}$. After receiving $\mathcal{I}^{'}$, system starts immediate contact tracing on $\mathcal{G}$ where $\mathcal{I}^{'}$, $\Gamma$, $\mathcal{G}$ and $L$ are passed as input parameters to Algorithm 1 (Appendix). Here, $L\geq 1$ represents up to which level we trace the contacts. The algorithm returns updated $\Gamma$, which contains first, second, and other required higher levels of indirect contacts along with respective edge list $\chi^{'}$. Finally, we update new infected list as $\mathcal{I} \leftarrow \mathcal{I} \cup \mathcal{I}^{'}$, and directed edge list to detect infection pathways as $\chi \leftarrow \chi \cup \chi^{'}$. 

The algorithm uses two queues ($Q_1$ and $Q_2$) data structure \cite{cormen2009introduction} alternatively as processing and waiting of individuals to trace close contacts (Algorithm 1 (Appendix)). Initially, we set the current contact level number $l=1$ (direct or first-level contact) and $Q_1$, $Q_2$ to be empty. Here, $\mathcal{G}.GetNext(P)$ function gives the next individual member along with the associated contact vector address of the adjacency list, including overflow area (if any). For each infected member in $\mathcal{I}^{'}$, from the adjacency list in $\mathcal{G}.\Psi$, we prepare the direct contact list. At the same time, we insert the adjacent information into the processing queue ($Q_1$) for the second level contact trace and use $TraceOperator(\cdot)$ method (Algorithm 3 (Appendix)). For higher-level contact tracing, we repetitively update $l$ and swap $Q_1$ and $Q_2$. The process will continue either no more individuals left in the waiting queue or the target level ($L$) of indirect contacts has been reached (Algorithm 1 (Appendix)). Importantly, to decide if there is any indirect contact or not, we use Algorithm 3 (Appendix), which operates on two CCQs between $(P, P^{'})$ and  $(P^{'}, P^{''})$ and returns a \textsc{true}/\textsc{false} decision (Lemmas 1 and 2, Appendix). 
\begin{figure*}[tbh]
\centering
\includegraphics[width=\linewidth,height=10cm]{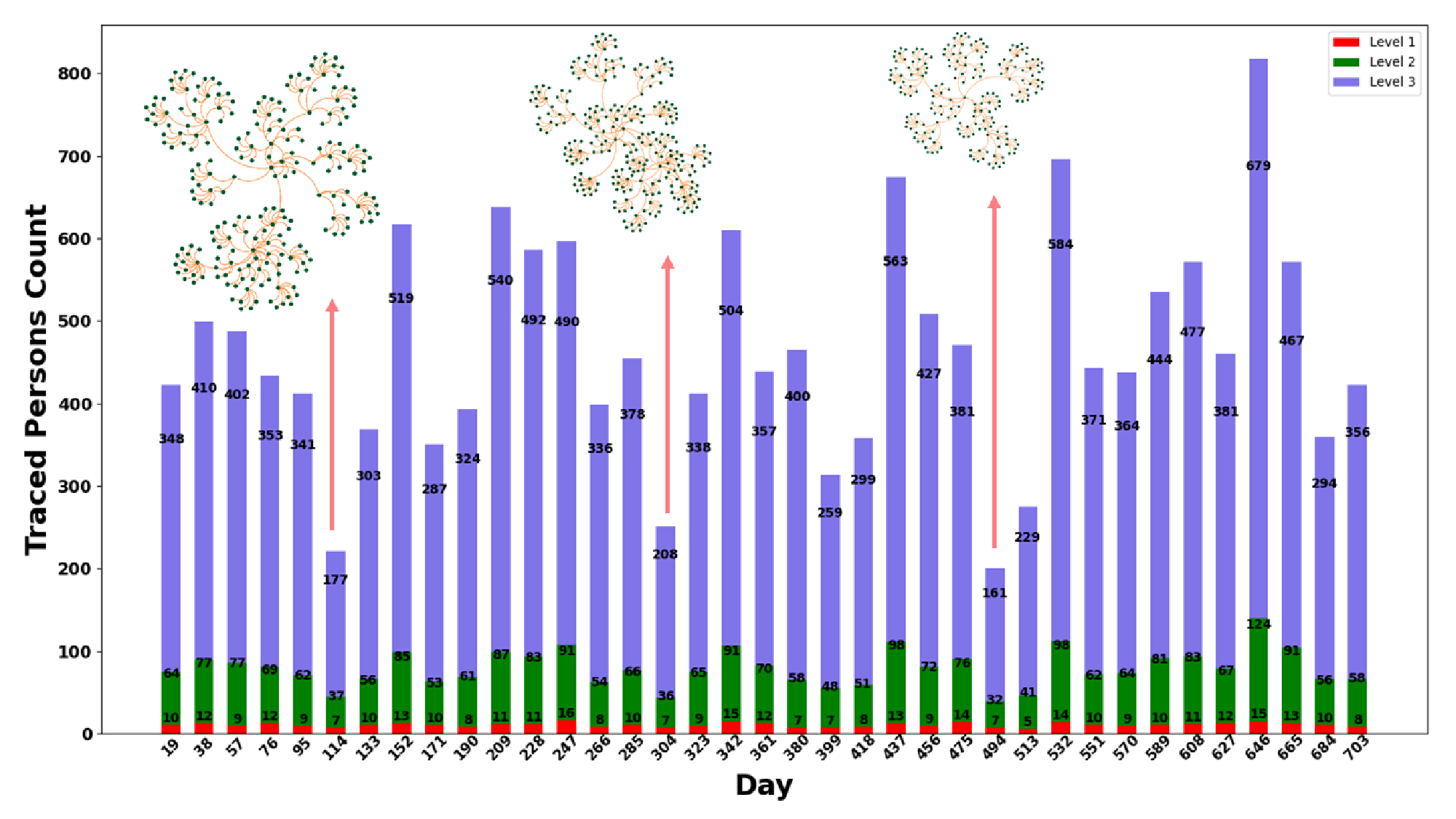}
    \caption{Depicting the counts of the traced persons for three levels for infected persons. Here, $N = 10^{5}$, $\langle q \rangle =10$, $D=10$, $L=3$ and $|\mathcal{I}^{'}|=1$ over two years. The algorithm counts the three-level traced persons for the randomly selected days and randomly selected infected persons. We mark first-level counts (red), second-level (green), and third-level (violet). For a random day, let's say on $114th$ day, a person is detected as infected by lab testing. Then, the contact trace algorithm finds the first-level contacts ($7$ persons), second-level contacts ($37$ persons), and third-level contacts ($177$ persons). The algorithm can identify the persons traced from the contact graph and create the infection pathways trees.}
\label{level_trace_100000}
\end{figure*}

\subsection{Synthetic Data}\label{synthetic_data}
To test the multilevel contact tracing algorithm, we require daily close contact data for a population. It is very difficult to get enough real-world close contact data to test the tracing algorithm. Therefore, we generate close contacts as synthetic data for our experiment. We fix a population of size $N$ for a particular zone. In a day, $N$ individuals can have at most $N(N-1)/2$ number of contacts leading to a complete graph. But in the real world, it cannot happen \cite{human_mobility_2010}; there will be a very small number of close contacts among the individuals. To capture it, we consider each day there are $m=\frac{\langle q \rangle N}{2D}$ number of close contacts, and that leads to $\langle q \rangle$ contacts on average for any day after completing the first incubation period ($D$). To generate contact data for a day, we randomly pick two individuals from the population and consider they have a close contact. We repeat the process by randomly choosing another pair of individuals up to $m$ pairs. The same process repeats for another day, and so on. This will create random independent day-wise close contact data leading to the Binomial degree distribution for the close contact graph (Fig. A6). Now, we apply the contact tracing algorithm on a toy example as described in Fig. \ref{contact_graph_representation}(a) and \ref{ExampleContactGraph}(Day 6). We consider $N=10$, $\langle q \rangle=2$, $D=5$ days, $\tau=1$ day, and $L=3$. We generate data for the six days and on the sixth day infected list contains one infected person as  $\mathcal{I}^{'}=\{P_2\}$. Hence, using Algorithm 1 (Appendix) we can find the multilevel contact trace list and infection pathways tree (Fig. \ref{contact_graph_representation}(c)). 

Next, we replicate the process for large contact datasets. We select a population size of $N=10^5$, with an average contact degree of $\langle q \rangle = 10$, a tracing level of $L=3$, an incubation period  $D=10$ days, and $\tau=1$ day. We generate day-wise close contact data for two years. For simplicity, we assume that people become infected randomly, and for a randomly chosen day, there is only one infected person i.e., $|\mathcal{I}^{'}|=1$. However, in reality, the number of lab-tested infected persons can vary significantly in a day. Appendix Figs. A7, A8 contains results for more than one infected person on a day.

After receiving the list of infected persons, our tracing algorithm will first identify all individuals who directly contacted the infected persons on a given day. It will then trace individuals who had contact with these traced persons, potentially being infected (indirect contact with an infected person). The algorithm will continue tracing up to the third level, expanding the search to capture a broader network of potentially susceptible individuals. The count of traced persons at specific levels for some random days of infected persons and the infection pathways tree are shown in Fig. \ref{level_trace_100000}. According to the contact graph, the average degree is $10$; in the first level, approximately $10$ people get infected, and from these ten people, another $10$, i.e., approximately $100$, and so on in the third level. Here, branching factor is $10$. 
%Note that once a person is detected as infected, the infected and contacted individuals in different level will be in the hospital or quarantine for a certain period. However, once they recover or the quarantine period is over, they may come into contact with others again, and the contact graph can have edges. The cycle repeats.  

\begin{figure*}[tbh]
\centering
\includegraphics[width=\linewidth,height=6cm]{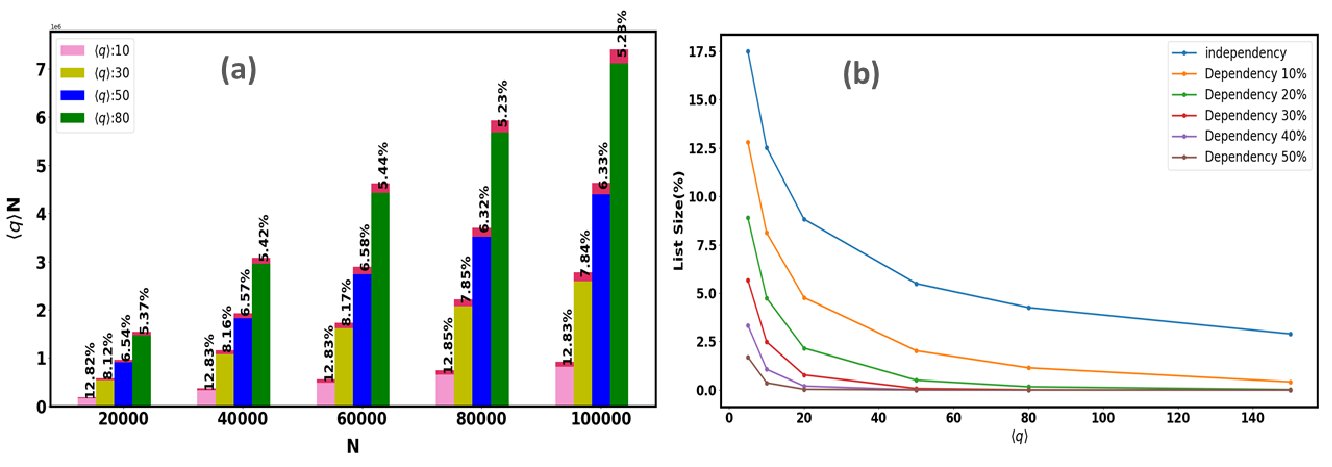}
\caption{(a) We vary population size ($N$) and average number of close contacts ($\langle q \rangle$) and portray the percentage of required space by link list. We performed the experiment for two years and observed the average link list size. For instance, for the fix $\langle q \rangle=10$, $N=10000$, the average link list size over the two-year duration is $11\%$ of the total space. We repeat the experiment for different $N$ and $\langle q \rangle$. (b) Further, we repeat the experiment when contact data is dependent and observe that increasing dependency decreases the space required by the link list.}
\label{linked_list_scale}
\end{figure*}

\subsection{Performance Analysis}\label{analysis}
Finally, we measure the space and time complexity of the contact tracing process.

\subsubsection{Space complexity analysis} \label{space_complx}
Here, we calculate the space requirement to store the contact graph sketch for the $D$ days proximity data of $N$ number of users as adjacency list representation. We know $\mathcal{G}$ has two parts, $\mathcal{G}.\Psi$ to store adjacent information and $\mathcal{G}.\Theta$, which stores circular contact queues and space requirement is denoted as $\mathcal{S}(\mathcal{G}.\Psi)$ and $\mathcal{S}(\mathcal{G}.\Theta)$ respectively. We need $b=\lceil log_2 N\rceil$ bits to access the UserID of the $N$ number of different users. Next, we assume, on average, $\langle q \rangle$ number of distinct close contacts for a days. We store $\langle q \rangle+1$ number of direct index records for each user in $\mathcal{G}.\Psi$. Each node has $\langle q \rangle $ average degree, hence total $\frac{\langle q \rangle N}{2}$ number of edges in $\mathcal{G}$. Therefore, $\mathcal{G}.\Theta$ will store $\frac{\langle q \rangle N}{2}$ CCQs each of size $n$ bits where $n=\lceil \frac{1440D}{\tau}\rceil$ number time slots of duration $\tau$ min. Further, to identify each such contact queue, we require $s=\lceil log_2 \frac{\langle q \rangle N}{2} \rceil$ bits in the address field of $\mathcal{G}.\Psi$. Therefore, the space requirement for $\mathcal{G}.\Psi$ is 
\begin{equation}\nonumber
\begin{split}
\mathcal{S}(\mathcal{G}.\Psi) &= (\langle q \rangle + 1)N(b+s) \\
&= (\langle q \rangle +1)N(2\log N + \log \langle q \rangle - 1) \\
    &=\mathcal{O}(N \log N)
\end{split}    
\end{equation}
Similarly, to store $\frac{\langle q \rangle N}{2}$ number of circular contact queues in $\mathcal{G}.\Theta$ require
\begin{equation}\nonumber
\mathcal{S}(\mathcal{G}.\Theta) = \frac{\langle q \rangle N}{2}n
=\mathcal{O}(N)
\end{equation}
Hence, the total space requirement (in bits) is 
\begin{equation}\label{space_com}
\begin{split}
\mathcal{S}(\mathcal{G}) &= \mathcal{S}(\mathcal{G}.\Psi) + \mathcal{S}(\mathcal{G}.\Theta) \\
     &= N\biggl[(\langle q \rangle+1)(2\log N + \log \langle q \rangle - 1)+\frac{n \langle q \rangle}{2}\biggr]\\ 
     &=\mathcal{O}(N \log N) 
\end{split}
\end{equation}
Here, $D$, $\tau$, and $\langle q \rangle$ are constant for a particular type of epidemic, and hence, $n$ is also constant, and we get the space complexity of $\mathcal{G}$ is $\mathcal{O}(N \log N)$. Although the fixed structure will take space $\mathcal{O}(N \log N)$, however when the degree of a node is below $\langle q \rangle$, some space will be empty in the array ($\mathcal{G}.\Psi$). Further, when the degree of a node exceeds $\langle q \rangle$,  extra space will be needed outside $\mathcal{G}.\Psi$ (i.e., in link list). But, always the average degree of the contact graph is  $\langle q \rangle$. Thus, the total degree of the contact graph will be $\langle q \rangle N$, which splits into two different places in the array ($\mathcal{G}.\Psi$) and in the link list. 

Now, we analyze the extra space requirement due to the link list. We performed the same experiment for two years and tracked the space required by the link list. To understand the impact of $N$ and $\langle q \rangle$ on the link-list size, we vary both parameters and investigate the link-list behavior. We can observe that for a fix $\langle q \rangle$ as $N$ is increasing, the link list size is growing linearly, and some percent of the total space ($N \langle q \rangle$) is going to the link list, and the rest is going to $\mathcal{G}.\Psi$ (Fig. \ref{linked_list_scale}(a)). Further, as $\langle q \rangle$ is varying, one can observe that the require space for the list is decreasing (Fig. \ref{linked_list_scale}(a)) and for $\langle q \rangle=50$, $5\%$ of the total space is going to the link list. Previously, we considered that day-wise data is independent of each other day, but in the real world, day-wise contact data might be dependent. We create the dependent data for our experiment by taking some percentage of contact data from the previous day. By increasing the overlap, we can decrease the list size (Fig. \ref{linked_list_scale}(b)). We can see that when the dependency is $10\%$ and $\langle q \rangle \sim 50$, $2\%$ of the total degrees are stored in the link list (Fig. \ref{linked_list_scale}(b)). 
We can very easily understand the extra space behavior and dependency in contacts from the CCQ of the contact graph. When the day-wise data is independent, there is very little chance that contact between two persons will repeat very frequently in the incubation period, which means there will be a single entry in the CCQ in $D$ days, and new contacts will be added daily. However, in the case of dependent data, a few contacts will repeat for the day, which means there is more than one entry in the CCQ, which leads to a decrease in the link list size. Therefore, on average, for the link list, extra space is taking $\mathcal{O}(N)$, and thus, the total space requirement for our algorithm is again $\mathcal{O}(N \log N)$.

\begin{figure*}[tbh]
\centering
\includegraphics[width=5.8in, height=3.8in]{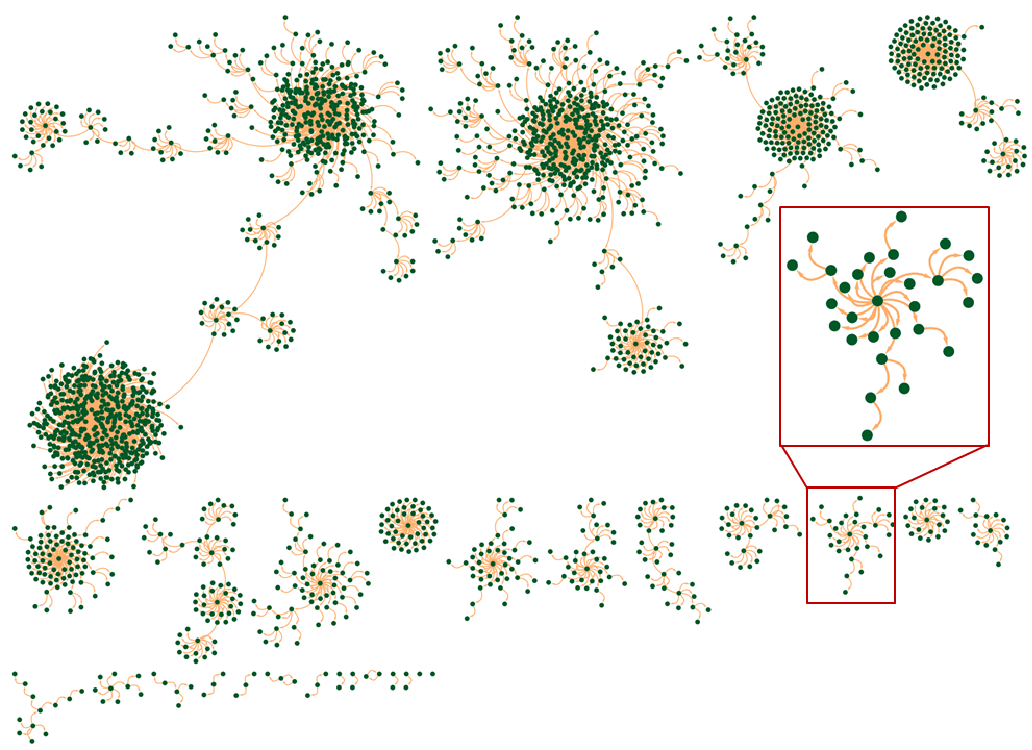}
\caption{Infection pathways for Korean contact tracing data from $23-01-2020$ to $04-03-2020$. Each node of the disjoint trees is the infected person.}
\label{Korean_data}
\end{figure*}

\subsubsection{Time complexity}\label{time_complx}
To detect a close contact in $\mathcal{G}$ requires a maximum $\langle q \rangle$ number of array index manipulations. We assume $\langle q \rangle$ is constant and independent of $N$; thus, computation complexity for the preparation of $\Gamma$ has two parts - computation counts for direct list preparation denoted as $(T_{direct}(\langle q \rangle, L,n, N))$ and for the indirect list as $(T_{indirect}(\langle q \rangle, L,n, N))$. Now, we evaluate $T_{direct}(\langle q \rangle,L,n,N)$ for given $\mathcal{I}^{'}$ as follows. For any infected person $P$ in $\mathcal{I}^{'}$, we access at most $\langle q \rangle$ number of consecutive pointer fields in $\mathcal{G}.\Psi$ array. We know that one can reach the beginning of the adjacency list of $P$ using $\mathcal{G}.\Psi[P\times(\langle q \rangle+1)]$ (sec. \ref{digitization_close_contact}). Therefore, total comparisons take
\begin{equation}\nonumber
T_{direct}(\langle q \rangle,L,n,N) =\langle q \rangle \times |\mathcal{I}^{'}| =\mathcal{O}(\langle q \rangle|\mathcal{I}^{'}|)
\end{equation}
However, to prepare the second level contact list, we perform $n$ bits contact queue comparison to execute $\sigma$ for each member in the direct contact list of $\langle q \rangle \times |\mathcal{I}^{'}|$ individuals (Algorithm 1 (Appendix)) is $\langle q \rangle\times (\langle q \rangle \times |\mathcal{I}^{'}|)$, and for third-level it is $q\times (\langle q \rangle^{2} \times |\mathcal{I}^{'}|)$ and so on. Hence, the total computation complexity for indirect levels is
\begin{equation}\nonumber
\begin{split}
T_{indirect}(\langle q \rangle,L,n,N)&= n[\langle q \rangle\times(\langle q \rangle\times|\mathcal{I}^{'}|) +\\
&\langle q \rangle\times(\langle q \rangle^{2}\times|\mathcal{I}^{'}|)
+\\
&\ldots+\langle q \rangle\times(\langle q \rangle^{L-1}\times|\mathcal{I}^{'}|)]\\
&=\langle q \rangle^{2}\times\frac{(\langle q \rangle^{L-1}-1)}{(\langle q \rangle-1)}\times (|\mathcal{I}^{'}|\times n)\\
&=\mathcal{O}(\langle q \rangle^{L}|\mathcal{I}^{'}|)
\end{split}
\end{equation}
Now, for both direct and indirect contact list tracing, computation takes 
\begin{equation}\label{time_complexity}
T(\langle q \rangle,L,n,N) =\mathcal{O}(\langle q \rangle^{L}|\mathcal{I}^{'}|)
\end{equation}
Finally, we use a small amount of real-world close contact data for the model verification.

\subsection{Real-world Contact Data: COVID-19}
For social, economic, and many other reasons, the inter-zones and inter-country transportation of humans are inevitable in real-world contact data. We collected two real-world COVID-19 data sets - South Korean and Indian data sets (Appendix Figs. A3 and A4). The South Korean contact data set for positive cases of the COVID-19 pandemic is available for $23-01-2020$ to $04-03-2020$ \cite{covid19_Korean_data}. Similarly, we use Indian positive cases patients data for the periods $30-1-2020$ to $10-04-2020$ \cite{covid19_Indian_data}. The collected data sets only contain the confirmed cases of the COVID-19 pandemic, complete contact trace data set with suspected, and others are not available. Further, for multiple instances, actual contact is unavailable or infected from a gathering event. Therefore, we carefully examine all such eventual contacts and preprocess the data sets accordingly.
After constructing the infection pathways for the South Korean data, we can observe that $N=2257$ individuals and $2245$ directed edges exist. Further, Indian data contains $N = 6734$ individuals and $6638$ directed edges (Appendix Fig. A5). These two countries have clustered transmission.

An important observation is that the outcome of the multilevel contact tracing algorithm provides infection pathways that are similar to the infection pathways received from the real-world Korean contact data sets (Figs. \ref{level_trace_100000},  \ref{Korean_data}). Furthermore, the real-world contact data only has information on infected persons and how others infect them. It needs to contain information on the complete close contacts.

Furthermore, we analyze the storage space and time required for our frameworks, and as a case study, we consider the COVID-19 outbreak. We assume for a particular zone, the number of population as $N=10^6$, the length of incubation period $D=14$ days, and minimum time of contact $\tau = 15$ min. To estimate $q$, we use statistics from a recent UK-based social contact survey \cite{danon2013social}. It suggests that for the COVID-19 pandemic, the average number of contacts ($q$) over $14$ days is $217$. However, only $59$ ($27\%$) of contacts are meeting close contact definition \cite{keeling2020efficacy,danon2013social}. With these statistics, we choose $q=64$. Hence, the estimated storage requirements for $\mathcal{G}$ using Eq. (\ref{space_com}) is: 
\begin{equation}\nonumber
\begin{split}
\mathcal{S}(\mathcal{G}) &=10^6[(64+1)(2\log 10^6 + \log 64 -1) \\
&+\biggl(\frac{1440\times 14}{15}\biggr)\frac{64}{2}]\; bits \approx 5\;GBytes
\end{split}
\end{equation}
Note that for simplicity, we consider $\langle q \rangle$ as a constant, which implies every node has the same degree. However, in real-world situations, $q$ follows a heterogeneous degree distribution \cite{network_sec_2016}. To handle this degree-heterogeneity in real-world implementations, we maintain an overflow area in $\mathcal{G}$ using a link list (Fig. \ref{contact_graph_representation}(b) and Fig. \ref{linked_list_scale}).  

Further, from Eq. (\ref{time_complexity}), we can also show that for COVID-19 contact trace parameters, to prepare the contact list for a given $\mathcal{I}^{'}$, it takes $\mathcal{O}(|\mathcal{I}^{'}|)$ when $q$ and $L$ are constant. 

\section{Discussion}\label{conclusion}
We develop a framework for a multilevel digital contact tracing system. Our work provides an algorithm for the multilevel digital contact tracing process, the data structure, and how we can implement it. Our algorithm receives discrete proximity-contact data, converts it into a fixed-length bit pattern, and stores it inside a circular contact queue (CCQ). Further, all CCQs form an efficient dynamic contact graph. Once we maintain the close contact graph, our algorithm prepares the direct and indirect (multilevel) contact list of the infected persons and creates the infection pathways trees. We generate synthetic data sets to test the contact tracing process and use some real-world contact data sets for validation. We provide analysis that ensures that the method will detect the contact trace list for a given infected individual.   
The binary representation of close contact is an integral part of our algorithm. The heart of our algorithm is the contact trace operator, which uses numerical computation over the binary number stored inside the circular contact queue to decide the multilevel contact trace list. Our analysis unveils that for COVID-19 outbreak close contact parameters, the storage space required to maintain the contact graph takes $\mathcal{O}(N\log N)$ bits, and extracting the contact trace list takes $\mathcal{O}(|\mathcal{I}|)$ time.

Importantly, interactions between a pair of individuals are annotated as a binary circular contact queue, which captures a complete temporal picture of the social interactions. Therefore, the contact graph reflects the population dynamics of a zone and can be used for further study related to predicting the trajectory of the epidemic dynamics spread \cite{oliver2020mobile}. Our framework is relevant to general contact-transmitted diseases and can trace digital contacts by changing the parameters of our algorithm (incubation period and proximity duration of contacts). On the other hand, we consider a country divided into zones and the mobility of users under one PHA, i.e., our system is centralized under a PHA. We can find the benefits of centralized systems over decentralized systems when considering the risk of identifying diagnosed people \cite{centralized_DCT_2020, centralized_DCT_2021, centralized_DCT_2022}. However, our model can be extended to distributed digital contact tracing systems where individuals can move among several PHAs nationwide. By considering the location and time-independent binary contact queues and making centralized control by the PHA, our algorithm maintains the individuals' privacy \cite{mello2020ethics,digital_contact_2023}. We consider a fixed $q$ number of array locations for the individual adjacency list to hold the $q$ number of contact data. The $(q+1)^{th}$ member of a particular list has a pointer pointing to the overflow area where we store other than $q$ contacts. For simplicity, we maintain a separate link list structure for these extra contact data in this overflow area. However, we can also use the AVL, Red-Black, B-Trees, or hashing \cite{cormen2009introduction} for both fixed size $q$ elements array and the records in the overflow area, which needs further investigation. 

We provide a test bed for multilevel digital contact tracing where one can generate synthetic contact data and test their models. The main objective of this Information Technology system is to make solutions feasible to computation that can control the further transmission of diseases. However, as in practical situations, we should follow other pre-existing processes like social distancing, partial lockdown, preventive measures like washing hands, and disinfection to control infections.

\ifCLASSOPTIONcompsoc
  % The Computer Society usually uses the plural form
%   \section*{Acknowledgments}
% \else

  % regular IEEE prefers the singular form
  \section*{Acknowledgment}
\fi
Gautam Mahapatra acknowledges the Post-graduate Division of Asutosh College, University of Calcutta, for financial support and computational resources. GM is thankful to Ranjan Chattaraj (BIT Mesra, India) and Soumya Banerjee (Inria EVA, France) for reading the article and for the valuable suggestions during the early stages of the work. Priodyuti Pradhan acknowledges the Science and Engineering Research Board (SERB) grant TAR/2022/000657, Govt. of India.
\ifCLASSOPTIONcaptionsoff
  \newpage
\fi

%%%-----------------------------
\newpage
\section*{Appendix}%\label{methods}
\section{Multilevel Tracing}
We can identify benefits of the multilevel contact tracing as -- 

\vspace{2mm}
\noindent {\bf a) Greater coverage.} It extends the reach of tracing efforts beyond just the immediate contacts of confirmed cases, ensuring a more thorough identification of potential transmission chains. 

\vspace{2mm}
\noindent {\bf b) Early detection.} By tracing contacts beyond the primary contacts, multilevel contact tracing can detect potential cases earlier in the transmission chain, allowing for prompt isolation and treatment, thereby preventing further spread. 

\vspace{2mm}
\noindent {\bf c) Early warning.} Identifying contacts beyond immediate interactions can provide early warning signs of potential outbreaks, allowing for timely intervention. 

\vspace{2mm}
\noindent {\bf d) Enhanced control.} Multilevel tracing generates rich data that can be used for detailed analysis and modeling of disease spread, leading to more informed decision-making. By tracing multiple levels of contact, public health authorities (PHA) gain a deeper understanding of transmission dynamics, allowing for more targeted and effective interventions. It also helps to optimize resource allocation by focusing on high-risk contacts and locations. 

\vspace{2mm}
\noindent Overall, multilevel contact tracing can significantly enhance the effectiveness of public health interventions in managing and controlling infectious diseases.

\section{Lemmas}
 
From the indirect contact tracing process (sec. 2.2.2), we can observe three different cases, which we can explain from the properties of the binary number system and discuss in the following lemmas.  

\vspace{0.5 \baselineskip}
\noindent {\bf Lemma 1.} Let $\bm{c}_{1}=\bm{c}_{(P,P^{'})}=c_{n-1}\ldots c_0$ and $\bm{c}_{2}=\bm{c}_{(P^{'},P^{''})}=c_{n-1}^{'}\ldots c_0^{'}$ are two aligned nonzero binary contact vectors among three individuals $P,P^{'}$ and $P^{''}$ in $\mathcal{G}$ where $c_{n-1}$ represents the oldest and $c_0$ represents the latest days. Also there is no edge between $P$ and $P^{''}$ in $\mathcal{G}$. Suppose $P$ is identified as infected, and $val(\bm{c}_1) \geq val(\bm{c}_2)$, then $P^{''}$ will be included in the indirect trace list of $P$. Here, $val(\cdot)$ returns the decimal value of the binary contact vector.

\vspace{2mm}
\noindent {\bf Proof.}  
For 
\begin{equation}\nonumber %\label{eq:2}
    val(\bm{c}_1) \geq val(\bm{c}_2)
\end{equation}
We have $val(\bm{c}_1)=val(\bm{c}_2)\neq 0$, and thus $c_i=c_i^{'},\;\;\forall i=0,\ldots n-1$, and $\exists i\; \text{such that}\; c_i=c_i^{'}=1$. This implies in the last $D$ days while $P$ is in close contact with $P^{'}$, then $P^{'}$ is also in close contact with $P^{''}$ on that day. Therefore, there is a possibility of infection transmission from $P$ to $P^{"}$ via $P^{'}$. Now, for $val(\bm{c}_1) > val(\bm{c}_2)$, we have  
\begin{equation}\label{lem_1}
\begin{split}
\bm{c}_{1}=\bm{c}_{(P,P^{'})} &=\underbrace{c_{n-1}c_{n-2} \ldots c_{n-m}}_{m}1c_{n-m-2} \ldots c_0\\
\bm{c}_{2}=\bm{c}_{(P^{'},P^{''})} &=\underbrace{c_{n-1}^{'} c_{n-2}^{'} \ldots c_{n-m}^{'}}_{m}0c_{n-m-2}^{'} \ldots c_{0}^{'}
\end{split}    
\end{equation}
We know that for any binary integer number, the weight of a higher significant position ($2^m$) is always greater than the sum of all weights at lower significant positions ($m-1$ to $0$) \cite{cormen2009introduction}. For this reason, if the decimal value of any binary number ($v_1 = val(\bm{c}_{(P, P^{'})})$) is greater than equal to the decimal value of another nonzero binary number ($v_2=val(\bm{c}_{(P^{'}, P^{"})})$), indicates that at least one close contact occurs earlier in the circular contact queue corresponds to $v_1$ than all close contacts corresponds to $v_2$. 

\vspace{0.5 \baselineskip}
\noindent {\bf Case I:} From Eq. (\ref{lem_1}), oldest or most significant $m$ ($0\leq m \leq n-1$) number of bits are zeros. Then we have
\begin{equation}\nonumber
\begin{split}
\bm{c}_1&=\underbrace{0 \ldots 0}_{m}1c_{n-m-2} \ldots c_0\\
\bm{c}_2&=\underbrace{0 \ldots 0}_{m}0c_{n-m-2}^{'} \ldots c_{0}^{'}
\end{split}    
\end{equation}
It indicates that oldest $m$ number of days there is no contact between ($P,P^{'}$) and ($P^{'},P^{''}$). On the $(m+1)th$ day, there is a contact between $P$ and $P^{'}$  and no contact between $P^{'}$ and $P^{''}$. However, as $\bm{c}_2$ is nonzero, there is at least one close contact between $P^{'}$ and $P^{"}$ during  $c_{n-m-2}^{'} \ldots c_{0}^{'}$ days. Therefore infection transmission from $P$ to $P^{''}$ via $P^{'}$ is always possible. 

\vspace{0.5 \baselineskip}
\noindent {\bf Case II:} Further from Eq. (\ref{lem_1}), the oldest or most significant $m$ ($0\leq m \leq n-1$) number of bits are the same (any combination of one and zeros) in $\bm{c}_1$ and $\bm{c}_2$. Then, we get 
\begin{equation}\nonumber
\begin{split}
\bm{c}_1 &=\underbrace{1 \ldots 0 \ldots 1}_{m}1c_{n-m-2} \ldots c_0\\
\bm{c}_2 &=\underbrace{1 \ldots 0 \ldots 1}_{m}0c_{n-m-2}^{'} \ldots c_{0}^{'}
\end{split}    
\end{equation}
In the above, within $m$ most significant bits, there exists any bit which is one for both the vectors. It indicates that on the same day, $P$, $P^{'}$, and  $P^{"}$ are close to each other, and infection can be transmitted. 

Hence, the infection can always transmit for $v_1 \geq v_2$. Importantly, if $i$ is the oldest bit position in $\bm{c}_1$, which is one, and $j$ is the latest bit position in $\bm{c}_2$, which is one. Then for $v_1 \geq v_2$, always $i \geq j$.

\vspace{0.5 \baselineskip}
\noindent {\bf Lemma 2.} 
Let $\bm{c}_{1}=\bm{c}_{(P,P^{'})}=c_{n-1}\ldots c_0$ and $\bm{c}_{2}=\bm{c}_{(P^{'},P^{''})}=c_{n-1}^{'}\ldots c_0^{'}$ are two aligned nonzero binary contact vectors among three individuals $P,P^{'}$ and $P^{''}$ in $\mathcal{G}$ where $c_{n-1}$ represents the oldest and $c_0$ represents the latest days. Also there is no edge between $P$ and $P^{''}$ in $\mathcal{G}$. Suppose $P$ is identified as infected, and $val(\bm{c}_1) < val(\bm{c}_2)$ and $i \geq j$, then $P^{''}$ will be included in the indirect trace list of $P$. Here, $i$ is the oldest bit position in $\bm{c}_1$, and $j$ is the latest bit position in $\bm{c}_2$ those are one.

\vspace{0.5 \baselineskip}
\noindent {\bf Proof.}
For $val(\bm{c}_1) < val(\bm{c}_2)$, we have  
\begin{equation}\label{lem_2}
\begin{split}
\bm{c}_1 &=\underbrace{c_{n-1}c_{n-2} \ldots c_{n-m}}_{m}0c_{n-m-2} \ldots c_0\\
\bm{c}_2 &=\underbrace{c_{n-1}^{'} c_{n-2}^{'} \ldots c_{n-m}^{'}}_{m}1c_{n-m-2}^{'} \ldots c_{0}^{'}
\end{split}    
\end{equation}

\vspace{0.5 \baselineskip}
\noindent {\bf Case I:} From Eq. (\ref{lem_2}), oldest or most significant $m$ ($0\leq m \leq n-1$) number of bits are zeros. Then we have
\begin{equation}\nonumber
\begin{split}
\bm{c}_1&=\underbrace{0 \ldots 0}_{m}0c_{n-m-2} \ldots c_0\\
\bm{c}_2&=\underbrace{0 \ldots 0}_{m}1c_{n-m-2}^{'} \ldots c_{0}^{'}
\end{split}    
\end{equation}
It indicates that oldest $m$ number of days there is no contact between ($P,P^{'}$) and ($P^{''},P^{'}$). On the $(m+1)th$ day, there is no contact between $P$ and $P^{'}$ and a contact between $P^{'}$ and $P^{''}$. However, as $\bm{c}_1$ is nonzero, so there is at least one close contact between $P$ and $P^{'}$ on $ith$ day during $c_{n-m-2} \ldots c_{0}$ days. Now if $j \in c_{n-m-2}^{'} \ldots c_{0}^{'}$ and all are zeros or $jth$ bit is one but $i<j$ then infection transmission is not possible. As $P^{'}$ and $P^{''}$ meet on $(m+1)th$ day and later $P$ and $P^{'}$ meet which is captured by $i<j$ with $val(\bm{c}_1) < val(\bm{c}_2)$. However, we emphasize that if there is any $jth$ bit which is one in $c_{n-m-2}^{'} \ldots c_{0}^{'}$ and $j$ is lesser than $i$ then infection transmission is possible. Although $val(\bm{c}_1) < val(\bm{c}_2)$, but $i \geq j$ says that $(P^{'},P^{''})$ meet after the $(P,P^{'})$, thus insfection can transmit from $P$ to $P^{''}$ via $P^{'}$. 

\vspace{0.5 \baselineskip}
\noindent {\bf Case II:} Further in Eq. (\ref{lem_2}), the oldest or most significant $m$ ($0\leq m \leq n-1$) number of bits are the same (any combination of one and zeros) in $\bm{c}_1$ and $\bm{c}_2$. Then, we get 
\begin{equation}\nonumber
\begin{split}
\bm{c}_1 &=\underbrace{1 \ldots 0 \ldots 1}_{m}0c_{n-m-2} \ldots c_0\\
\bm{c}_2 &=\underbrace{1 \ldots 0 \ldots 1}_{m}1c_{n-m-2}^{'} \ldots c_{0}^{'}
\end{split}    
\end{equation}
In the above, within $m$ most significant bits, there exists any bit which is one for both the vectors. It indicates that on the same day, $P$, $P^{'}$, and  $P^{"}$ are close to each other, and infection can be transmitted. Here, $v_1 < v_2$ but $i \geq j$. Hence, for $v_1 < v_2$ and $i \geq j$, infection can transmit from $P$ to $P^{''}$ via $P^{'}$.

\begin{algorithm}
\caption{TraceContacts($\mathcal{G}, \mathcal{I}^{'}, \Gamma, L$)}
\label{ContactTracing}
\begin{algorithmic}[1]
\STATE $\chi=\{\}, Q_1=\phi, Q_2=\phi, l=1$
\FORALL{$P \in \mathcal{I}^{'}$}
\STATE $P^{'},\alpha \leftarrow \mathcal{G}.GetNext(P,\FALSE)$
\WHILE{$ P^{'}\geq 0 \; \AND \;\alpha \neq NULL $}
%\STATE $P^{'} \leftarrow \mathcal{G}.\Psi[\alpha].UID$
\IF{$P^{'}\notin \Gamma $}
\STATE $P^{'}.Level \leftarrow l,\Gamma \leftarrow \Gamma \cup P^{'},\chi \leftarrow \chi \cup (P, P^{'})$
\STATE $Q_1 \leftarrow Insert(Q_1, (P^{'},\alpha))$
\ENDIF
\STATE $P^{'},\alpha \leftarrow \mathcal{G}.GetNext(P,\TRUE)$
\ENDWHILE
\ENDFOR
\STATE $l \leftarrow l + 1$
\WHILE {(($l \leq L$) \AND \NOT ($IsEmpty(Q_1)$ \AND $IsEmpty(Q_2)$))}
\WHILE {(\NOT $IsEmpty(Q_1)$)}
\STATE $Q_1 \leftarrow Delete(Q_1,(P,\alpha)),\;\bm{c}_1 \leftarrow \mathcal{G}.\Theta[\mathcal{G}.\Psi[\alpha].Adress]$ 
%\STATE $\beta \leftarrow P \times (q+1)$
\STATE $P^{'},\beta \leftarrow \mathcal{G}.GetNext(P,\FALSE)$
\WHILE{$P^{'}\geq 0 \;\AND \;\beta \neq NULL $}
\STATE $\bm{c}_2 \leftarrow \mathcal{G}.\Theta[\mathcal{G}.\Psi[\beta].Adress]$
\IF{$P^{'} \notin (\Gamma \cup \mathcal{{I}^{'}} )\; \AND\; TraceOperator(\bm{c}_1,\bm{c}_2)$}
\STATE $P^{'}.Level \leftarrow l,\Gamma \leftarrow \Gamma \cup P^{'},\chi \leftarrow \chi \cup (P, P^{'})$
\STATE $Q_2 \leftarrow Insert(Q_2, (P^{'},\beta))$
\ENDIF
\STATE $P^{'},\beta \leftarrow \mathcal{G}.GetNext(P,\TRUE)$
\ENDWHILE
\ENDWHILE
\STATE $Q_1, Q_2 \longleftrightarrow Q_2,Q_1 $
\STATE $l \leftarrow l + 1$
\ENDWHILE
\RETURN $\Gamma, \chi$
\end{algorithmic}
\end{algorithm}

\begin{algorithm}
\caption{GetNext($P, Flag$)}
\label{GetNext}
\begin{algorithmic}[1]
\IF{$P <0 $}
\RETURN $-1, NULL$
\ENDIF
%\STATE [here $\alpha$ is a static type local variable]
\IF{$\NOT Flag $}
\STATE $\alpha \leftarrow P \times (q+1)$
\ENDIF
\STATE $P^{'} \leftarrow \mathcal{G}.\Psi[\alpha].UID$
\STATE $\beta \leftarrow \mathcal{G}.\Psi[\alpha].Adress$ 
\STATE $\alpha \leftarrow \alpha + 1$
\RETURN $P^{'}, \beta$
\end{algorithmic}
\end{algorithm}

\begin{algorithm}
\caption{TraceOperator($\bm{c}_1,\bm{c}_2$)}
\label{ImplementSigma}
\begin{algorithmic}[1]
\STATE $v_1 \leftarrow val(\bm{c}_1)$
\STATE $v_2 \leftarrow val(\bm{c}_2)$
\IF{$v_1 \geq v_2$}
\RETURN \textsc{true}
\ELSE  %% Added
\STATE $i \leftarrow oldest\;\;close\;\;contact\;\;of\;\;\bm{c}_1$
\STATE $j \leftarrow latest\;\;close\;\;contact\;\;of\;\;\bm{c}_2$
\IF {$i \geq j$}
\RETURN \textsc{true}
\ELSE
\RETURN \textsc{false}
\ENDIF
\ENDIF  %% Added
\end{algorithmic}
\end{algorithm}

\begin{table*}[tbh]
\begin{center}
\begin{tabular}{|l|p{4cm}|p{4cm}|p{4cm}|} 
\hline
 & {\bf Symptomatic}&{\bf Asymptomatic} &{\bf Presymptomatic}\\
\hline
{\bf Symptoms} &Symptomatic individuals develop noticeable symptoms of the disease &Asymptomatic individuals do not develop symptoms of the disease throughout the course of the infection.&Presymptomatic individuals have been infected but have not yet developed symptoms.\\ 
\hline
{\bf Timing} &Symptomatic individuals are infectious and can spread the disease to others while they are experiencing symptoms.&Asymptomatic individuals are infectious and can spread the disease to others even though they do not have symptoms. &Presymptomatic individuals are infectious before they develop symptoms, typically during the incubation period of the disease.
\\
\hline
{\bf Detection} &Symptomatic cases are typically easier to detect, as individuals seek medical attention when they develop symptoms. &Asymptomatic cases are often detected through testing of individuals who have been in close contact with confirmed cases or through surveillance testing.&Presymptomatic cases can be challenging to detect, as individuals are not yet showing symptoms. Contact tracing and testing of close contacts are essential for identifying presymptomatic cases.\\ 
\hline
{\bf Example} & In the context of COVID-19, symptomatic individuals experience symptoms such as fever, cough, and fatigue, and they can spread the virus through respiratory droplets. &Some individuals infected with the flu virus may not develop symptoms but can still spread the virus to others.&In the context of COVID-19, individuals infected with the virus can spread it to others in the days before they start showing symptoms.\\
\hline
\end{tabular}
\end{center}
\caption{Nature of Disease Spread \cite{asymptomatic_2020}.}
\label{disease_spread_nature}
\end{table*}

%\newcommand{\beginsupplement}{%
%\setcounter{equation}{0}
%        \renewcommand{\theequation}{S\arabic{equation}}
 %       \setcounter{table}{0}
  %      \renewcommand{\thetable}{S\arabic{table}}%
  %      \setcounter{figure}{0}
   %     \renewcommand{\thefigure}{S\arabic{figure}}%
    % }

%\section*{Introduction}
%\beginsupplement

%\begin{table*}[t]
%\begin{center}
%\begin{tabular}{|l|p{12cm}|} 
%\hline
%Symbols and Notations & Description \\
%\hline
%{$\tau, D$} & Close contact parameters -- minimum time duration spend for close contact (proximity duration) and latest observation period in days (incubation period)\\ 
%\hline
%$\bm{c}_{(P,P^{'})}$ & Contact vector -- between individuals $P$ and $P^{'}$. 
%\\
%\hline
%$\sigma $ & Contact trace operator -- a binary operator for indirect (multi-level) contact tracing. \\ 
%\hline
%$P \rightsquigarrow P^{'}$ & flow of infection elements from $P$ to $P^{'}$\\
%\hline
%$\mathcal{G}$ & Close contact graph -- having two components (i) ($\mathcal{G}.\Phi$) stores addresses of the close contact vectors (ii) $\mathcal{G}.\Theta$ stores close contact vectors.\\
%\hline
%$V$, $\mathcal{I}$, $\Gamma$, and $\chi$ & population having smartphones, infected persons, suspected individuals or contact trace list, directed edges for infection transmission pathways \\ 
%\hline
%\end{tabular}
%\end{center}
%\caption{List of symbols and notations.}
%\end{table*}

\begin{figure*}[ht]
\begin{center}
\includegraphics[width=5.8in, height=4in]{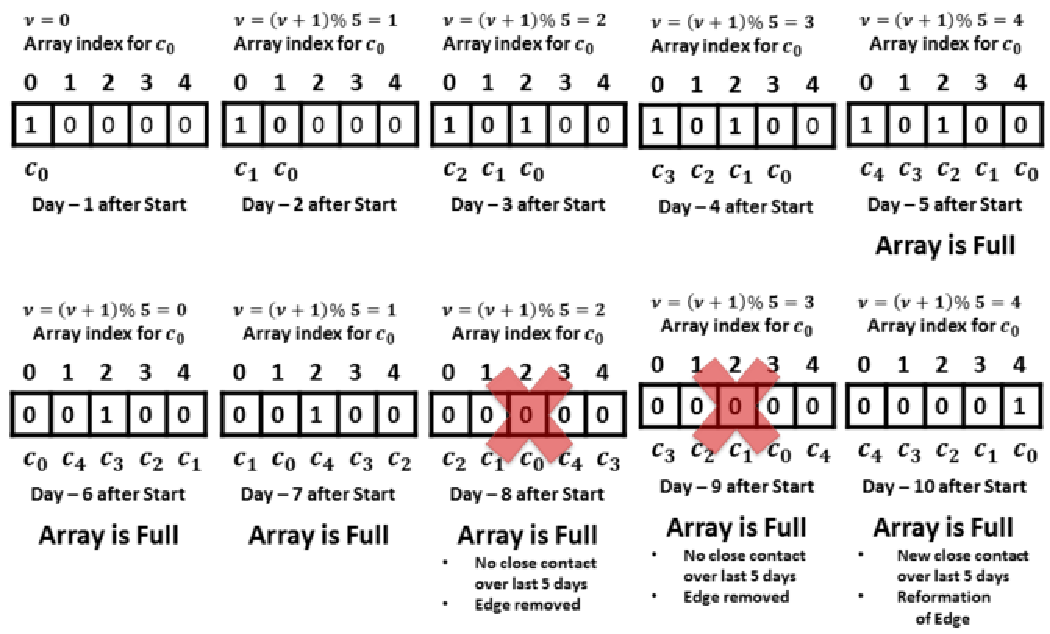}
\caption{Illustration of the array-based circular queue implementation using modular arithmetic for a contact vector $\bm{c}_{(P, P^{'})}$ in $\mathcal{G}$. For simplicity, we consider $D=5$ days, $\tau=1$ day, so the size of the contact vector is $n=5$. After $D$ days, $\bm{c}_{(P,P^{'})}$ attends the full length. Here, $\bm{c}=\bm{0}$ means there is no close contact during the latest $D$ days, so no infection transmission, and hence respective edge will be removed. However, they may come in close contact again.} \label{circular_queues}
\end{center}
\end{figure*}
\begin{figure*}[tbh]
    \centering
\includegraphics[width=6.2in, height=4.2in]{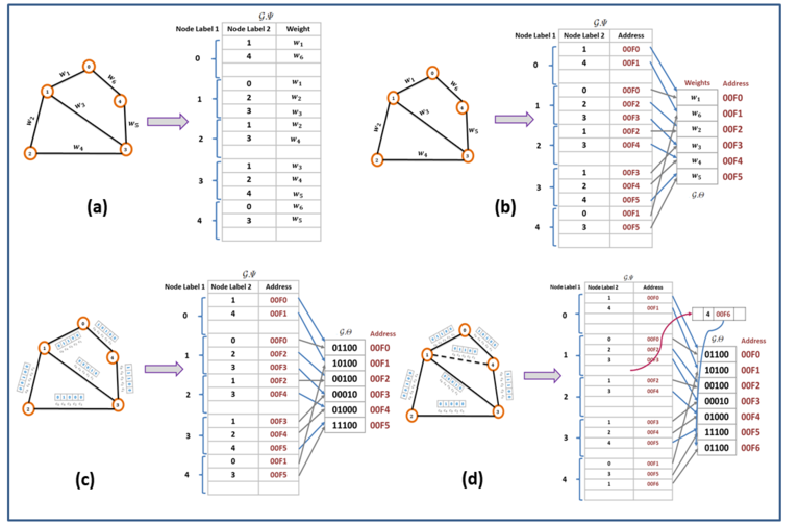}
\caption{Adjacency list graph representation. (a) sparse weighted graph representation with an array of weights associated with the edges. (b) sparse weighted graph representation with an array where weights are stored in different memory locations (c) Contact graph representation where contact queues are stored in separate memory locations. (d) The contact graph representation where double link list data structures implement the overflow area.}
\label{contact_graph_representation}
\end{figure*}

\begin{figure*}[h]
\centering
\includegraphics[width=1\linewidth]{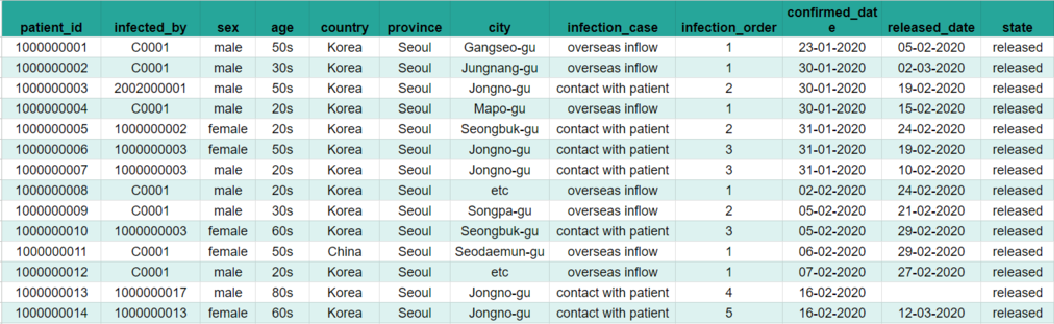}
\caption{Korean Patients Data from $23-01-2020$ to $04-03-2020$.}
\label{KoreaPatients}
\end{figure*}

\begin{figure*}[h]
\centering
\includegraphics[width=1\linewidth]{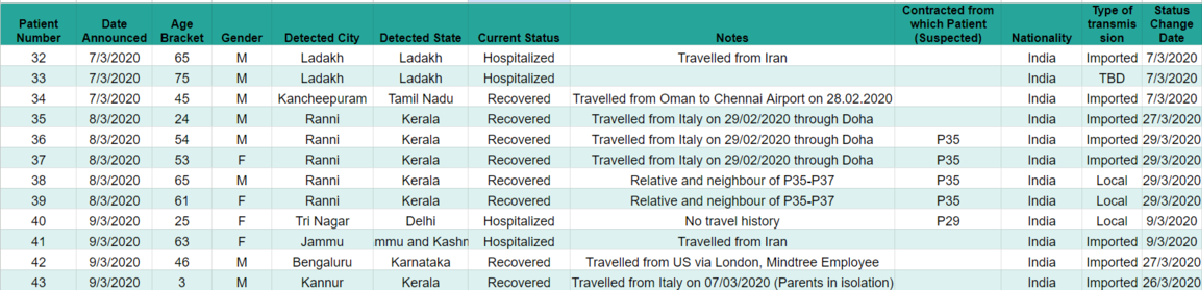}
\caption{Indian Patients Data from $30-1-2020$ to $10-04-2020$ on daily basis.}
\label{IndianPatients}
\end{figure*}

\begin{figure*}[tbh]
\centering
\includegraphics[width=6.2in, height=4in]{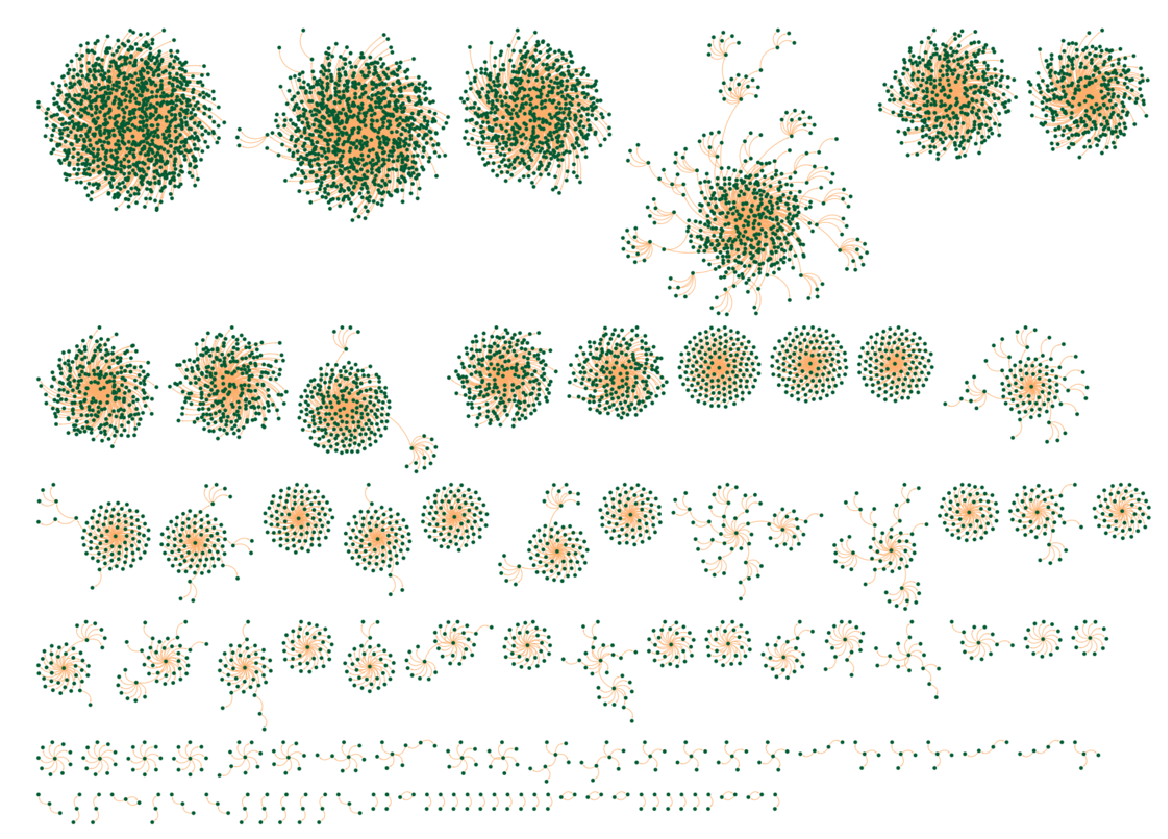}
\caption{Infection pathways of Indian contact tracing data with $6734$ individuals and contacts $6638$ contacts from $30-1-2020$ to $10-04-2020$.}
\label{Indian_contact_data}
\end{figure*}

\section{Queue and Circular Queue Data Structures}

A queue is a linear data structure that follows the First In, First Out (FIFO) principle \cite{cormen2009introduction}. It is similar to a queue of people waiting in line, the first person to join the queue is the first one to be served or processed, and new elements are added to the back while existing elements are removed from the front. 

In practice, a queue is also known as a buffer. Concerning the capacity, this buffer can be a limited size bounded buffer or mathematically infinite size unbounded buffer. In our case, it is act as bounded buffer. We use circular implementation of this buffer to efficiently utilize the assigned limited memory space without any movements of the queue members. Here, the server at the front itself is moving to the next element to serve; for this reason, existing members are not shifting to capture the deleted member's space.
Similarly, a new member is added in the empty place at the rear, which is available upon deleting a member from the front. So we can avoid all movements of the members in the queue. When the queue is full, a new member is placed at the same place of the current front, so without any data movement, the oldest member at the front is automatically getting out to accommodate the new member at the rear. Therefore, only the time-dependent latest members are present in the queue structure. However, all other non-circular queue implementations suffer either utilization of allocated memory space problem, requires movements of all existing queue members, or both \cite{cormen2009introduction}.

A circular queue is a data structure used in computer science and queuing theory \cite{cormen2009introduction}. It's a type of queue that operates in a circular manner, where elements are added and removed in a circular fashion. In a circular queue:

\noindent {\bf Circular Structure.} Instead of a linear structure where elements are added at one end and removed from the other, a circular queue wraps around, meaning that when you reach the end of the queue, the next element is added at the beginning.

\noindent {\bf FIFO (First-In-First-Out).} Like a traditional queue, the circular contact queue follows the FIFO principle, meaning the element that is inserted first is the one that gets removed first.

\noindent {\bf Efficient Use of Space.} Circular queues can efficiently use the available space in memory because, unlike a linear queue, when elements are removed, the space they occupied becomes available for new elements.

\noindent {\bf Implementation.} It's often implemented using an array or a linked list where the rear and front of the queue are connected to form a circle. The rear indicates the position where a new element can be inserted, while the front indicates the position of the first element in the queue.

Circular queues are commonly used in scenarios where there is a fixed amount of space allocated for a queue and the elements need to be cycled through in a continuous manner. They are utilized in various computer science applications like operating systems for managing resources, in networking for managing packets, and in simulations for event handling in operation research. We use circular queue to hold the incubation period temporal close contact data and termed as circular contact queue (Fig. \ref{circular_queues}).

%\subsection{Circular Contact Queue (CCQ)}

Importantly, as time advances to the first slot of $(D+1)^{th}$ day, the oldest slot at $c_{n-1}$ exits from the FIFO queue to accommodate a new incoming slot at $c_{0}$ (Fig. \ref{circular_queues}). For this queue stored in a fixed $n$ elements one dimension array structure with array indices $0,1,\ldots, n-1$  this entry and exit can be done in two different ways - (i) by keeping one end of the array as fixed, say lower array index $0$ for insertion or rear and other moving end towards higher array index for deletion or front (vice versa), for any insertion we do one array index shift of slot data towards higher index from lower index i.e. $c_{i+1} \leftarrow c_{i}$ for all array indices $i=0,1,\ldots, n-2$, and for full queue,  this replaces $c_{n-1}$ by $c_{n-2}$, and oldest slot $c_{n-1}$ gets out at highest array index location $(n-1)$, newest slot $c_{0}$ enters at lowest array index location $(0)$, but this is costly one as it needs $n-1$ number of physical shifts for slots data, however here the contact vector slot index and array index are equal, i.e. slot $c_{i}$ appears at array index $i$,  (ii) to avoid this shifting none of the insertion and deletion is restricted, i.e. both are moving, and hence any slot in the contact vector can be placed in any array index location, with this flexibility we use the circular or repetition properties of the integer array index in the closed range $[0,n-1]$, here beyond last array index i.e. beyond $n-1$ for one step ahead of the array access the cycling or repeatation starts from $0$, by this property when queue is full, i.e. all $n$ elements (or latest $D$ days) records are in it, then newer slot $c_{0}$ and oldest slot $c_{n-1}$ are at only one array index difference and hence incoming newest slot data due to $(D+1)^{th}$ day start can easily be placed at the array index of $c_{n-1}$, which is becoming newest slot $c_{0}$, earlier $c_{0}$ is becoming $c_{1}$, similarly without any physical movement rest all others upto $c_{n-2}$ slots are automatically one slot older i.e. $c_{i}$ is becoming $c_{i+1}$ for all $i=0,1,\ldots, n-2$; therefore with constant number of operations, without shifting, this circular form of FIFO queue stores latest $D$ days data. So $c_{0}$ and all other slot data can appear at any array index location, i.e., slot index and array index are not equal. Obviously, this circular contact queue is our choice. With these updated edge labels, our contract graph evolves and maintains the population's latest contact information.

\section{Security and Privacy}
For our system, the contact graph  ($\mathcal{G}$) holds individuals' contact data in a binary encoded form (contact vector - $\bm{c}$). This encoded contact data is a location-independent data model of the whole social contact structure ($V$) under PHA. We can easily protect the graph data from corruption using suitable private key cryptography. Locally captured discrete contact data also do not store any location data and the actual contact time. Our system only uses samples of the proximity-based presence of two low-power Bluetooth-enabled mobile phones and captures them as locally stored discrete contacts. It uses randomly selected virtual IDs from a set of virtual IDs assigned by the PHA and is updated regularly for privacy reasons \cite{rivest2020pact}. With these brief descriptive views, we can justify the system's reliability after maintaining security and privacy measures \cite{bengio2020need, mello2020ethics}.

\section{Real-World Contact Tracing Data sets}
The recent COVID-19 Pandemic outbreak has happened due to the spreading of different variants of the SARS-CoV-2 RNA virus through human social proximity close contacts. Several newly introduced digital contact tracing systems and long-existing manual contact tracing are playing crucial roles in mitigating the problem. However, very little real-world data concerning this contact tracing is publicly available \cite{trivedi2020digital, world2020digital,digital_contact_2023}.

Still, during the COVID-19 Pandemic, we have collected the South Korean contact data for positive cases daily from $23-01-2020$ to $04-03-2020$ \cite{covid19_Korean_data}. The 'patient-id' and 'infected-by' columns define transmission relation (Fig. \ref{KoreaPatients}). However, it is incomplete because, in several cases, there is no data in the 'infected-by' column, possibly due to infections from hospitals, churches, shopping malls, or others. In the preprocessing step, we assign unique codes for import, hospital, and church cases as various set members for the disjoint sets like other individual victims.

For India, we collected the positive cases of COVID-19 patients data from $30-1-2020$ to $10-04-2020$ daily \cite{covid19_Indian_data}. Here, the equivalent 'infected-by' attribute requires more preprocessing because, in several situations, even a brief description of the source is unavailable. We can observe the infection pathways in Fig. \ref{Indian_contact_data} where the number of individuals is $N = 6734$, and contacts are $E = 6638$. These two countries have clustered transmission. Note that in the data sets (Figs. \ref{KoreaPatients} and \ref{IndianPatients}), other data columns like age, gender, the status of the patients, address-related data, order of infection spreading, etc. may be used for many additional possible analysis for different other decisions.

\section{Multi-Precision Arithmetic}
Multi-precision arithmetic operations are needed to evaluate the contract trace operator \cite{cormen2009introduction}. Here, we use the following method for this arithmetic operation.
Assumptions:
\begin{enumerate}
    \item We consider the contact vector $\bm{c}$ of $n$ bits as an unsigned integer number for numeric operation. So, we can perform only the integer arithmetic operations on contact vectors.
    
    \item Implementation in programming language $(Lang)$ uses $w$ number of bits to represent the biggest integer data type. $w$ is divisible by $8$ and stored in a byte-organized primary memory system. We call them as big integers. For example, in C/C++, we have a 'long int' of $64$ bits.

    \item In practical situations, the size of the contact vector ($n$) is always larger than the size of the big integer data type ($w$) available in $Lang$. So, to process any integer arithmetic operation on a contact vector, like a numeric comparison of two contact vectors, we perform the same big integer numeric comparison multiple times, sequentially, starting from the MSB position of these two numeric forms of the contract vectors. For this reason, we call it multi-precision arithmetic operations.
    
    \item $n = k \times w + r$, that means when $n$ is not an exact multiple of $w$, we pad $r$ number of zero value bits at the MSB position to make the adjusted size as multiple of $w$.

    \item We store these $n+r$ bits in the byte-organized main memory system. Again, the number $n+r$ may not be divisible by $8$, and then some more zero-value bits are padded at MSB. Finally, the adjusted contact vector is stored in the main memory. The starting address ($Loc$) of this memory block points to a lower significant byte for the same.
\end{enumerate}

Arithmetic Operations in connection to the contact trace operator:
\begin{enumerate}
    \item Non-zero check (or zero check): Let us assume that $Loc$ is the main memory starting address of the contract vector we want to process. Starting from the most significant byte, we fetch $w$ number of bits as a big integer number ($a$). If $a$ is non-zero, then the contract vector $\bm{c}$ is not zero. Otherwise, we continue with the following lower significant $w$ number of bits. The process stops with non-zero or equal to zero when all bits are zero.
    
    \item Magnitude comparison of two non-zero contact vectors  $(\bm{c}_{1}, \bm{c}_{2})$: Let us assume that $Loc_{1}$ and $Loc_{2}$ are the starting address of $\bm{c}_{1}$, and $\bm{c}_{2}$ respectively. For the magnitude comparison, we start from the most significant byte location, which is starting at 
    $Loc_{i} + \lceil (n+r)/8 \rceil -1$ for $i=1, 2$. From these locations, $w$ bits of $\bm{c}_1$ and $\bm{c}_2$ are fetched as two big integer numbers stored in $a$ and $b$, respectively. If these two numbers are non-zero and unequal, a larger one implies that the corresponding contact vector is greater than the other. Similarly, a lesser one implies that the corresponding contact vector is smaller than the other. If these two numbers are zero or equal, we must consider the next $w$ number of bits for $a$ and $b$ and use the above comparison. A similar process will continue until we get one as greater than the other or finished with that $\bm{c}_1$ and $\bm{c}_2$ are non-zero equal.
    
    \item Oldest contact in a non-zero contact vector ($\bm{c}$): Let us assume that $Loc$ is the starting address of a given non-zero contract vector $\bm{c}$. Starting from most significant byte ($Loc + \lceil \frac{(n+r)}{8} \rceil-1$) 
    we fetch $w$ number of bits as big integer number stored in $a$. 
    If zero, proceed with the next lower significant $w$ number of bits as a big integer number $a$. The process continues until we get non-zero $a$. We are now using masking, starting from the most significant bit position of $a$; it is possible to identify the earliest contact in the non-zero contact vector ($\bm{c}$).
    
    \item Latest contact in a non-zero contact vector ($\bm{c}$): Let us assume that $Loc$ is the starting address of the given non-zero contract vector $\bm{c}$. Starting from the least significant byte ($Loc$), we fetch $w$ number of bits as a big integer number stored in $a$. If the value in $a$ is zero, proceed with the next lower significant $w$ number of bits as a big integer number $a$. The process continues until we get non-zero $a$. We are now using masking, starting from the most significant bit position of $a$; it is possible to identify the earliest contact in the non-zero contact vector ($\bm{c}$).   
\end{enumerate}

\section{Linked List as Dynamic space}

To efficiently manage contacts, we allocate an array of size equal to the average degree to each user. This array serves as a storage mechanism for contacts. However, for individuals with a higher degree of connectivity than the average, we utilize a linked list data structure.

\textbf{Reasons for Using Linked Lists}
Linked lists offer several advantages over other data structures for storing contacts:

\begin{itemize}
    \item \textbf{Efficient Addition}: Adding a new contact to a linked list can be done in constant time (\(O(1)\)). This is because new contacts can be inserted at the front of the linked list without the need for shifting existing elements.
    
    \item \textbf{Efficient Removal}: If we need to move a contact from the linked list to the array (e.g., when an index in the array becomes available), we can do so in constant time (\(O(1)\)). This is again because we can remove the contact from the front of the linked list without the need for shifting elements.
    
    \item \textbf{Contact Checking}: When determining whether any contact is with an infected person, we need to traverse the linked list. While this operation takes linear time (\(O(n)\)), it is still efficient. Additionally, alternative data structures would also require linear time or worse for this operation.
\end{itemize}

Thus, utilizing linked lists for storing contacts in digital contact tracing systems proves beneficial due to their efficient addition, removal, and contact checking operations, especially for individuals with high connectivity.

\begin{figure}[tbh]
\centering
\includegraphics[width=8cm,height=6cm]{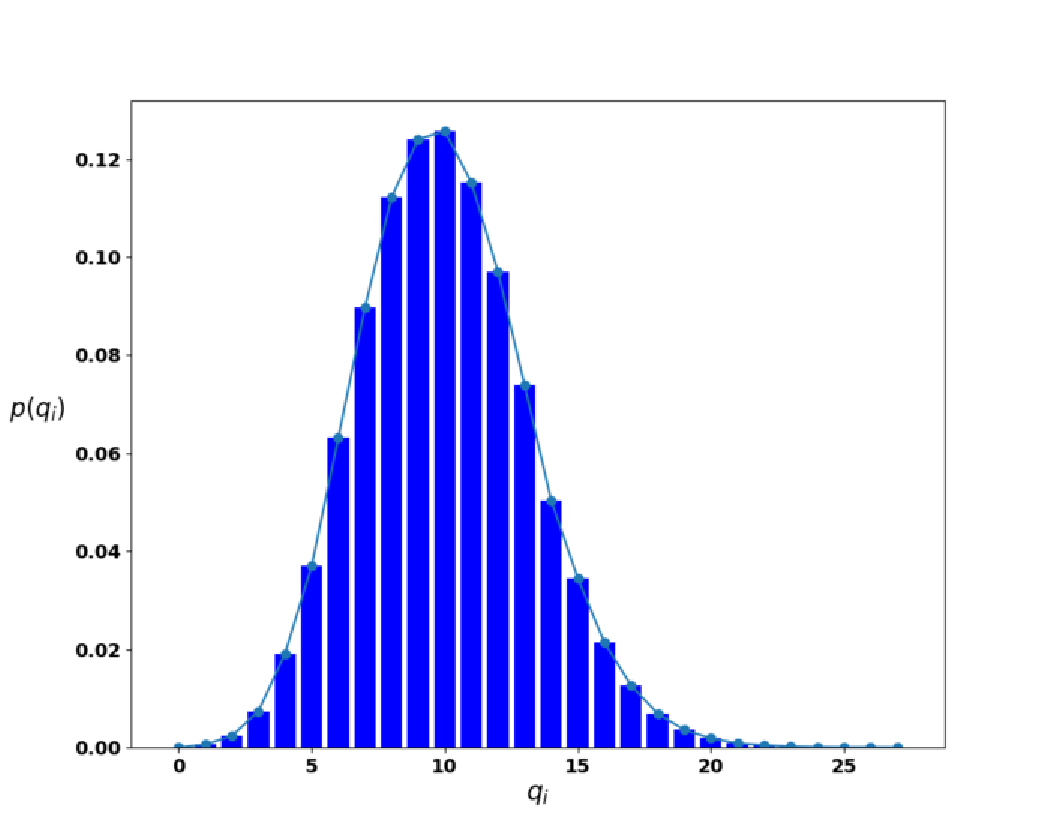}
    \caption{Degree distribution for a randomly chosen contact graph where population size  $N=100000$, $\langle q \rangle =10$.}
\label{contact_graph_deg_dist}
\end{figure}

\begin{figure*}[tbh]
\centering
\includegraphics[width=7.4in, height=4.2in]
{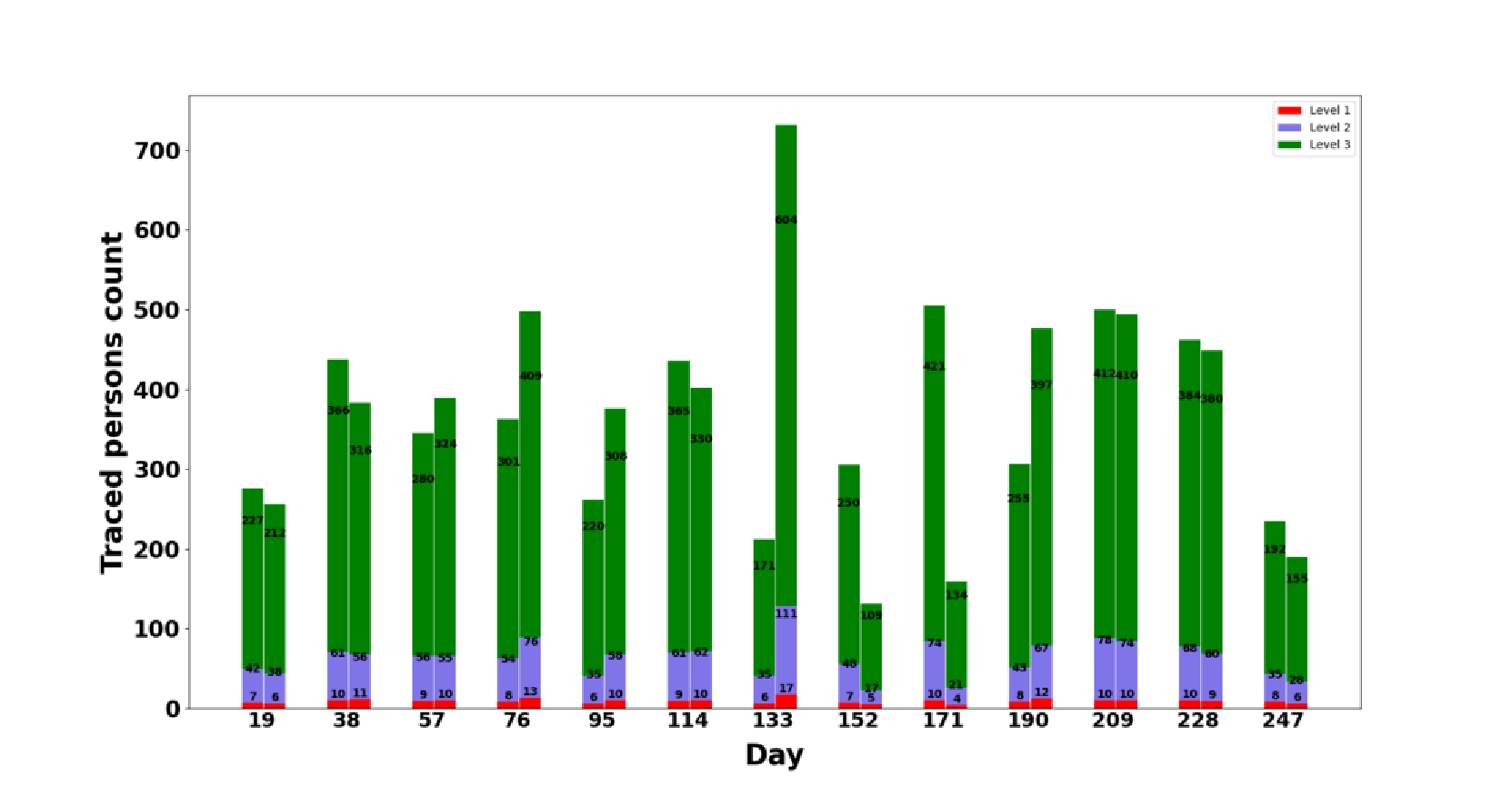}
\caption{N=100000, 300 days, each day two infected persons.}
\label{Indian_contact_data}
\end{figure*}

%\begin{figure*}[tbh]
%\centering
%\includegraphics[width=7.4in, height=4.2in]
%{images/plot_level_traced_3_infected.png}
%\caption{N=100000, 300 days, each day three infected persons.}
%\label{Indian_contact_data}
%\end{figure*}

% use section* for acknowledgment
\ifCLASSOPTIONcompsoc

\end{document}